\documentclass[twocolumn]{aastex63}

\usepackage{lineno}

\usepackage{graphicx}
\usepackage{xcolor}
\usepackage{epstopdf}
\usepackage{amsmath}
\usepackage{hyperref}
\usepackage{cleveref}
\usepackage{natbib}
\usepackage{CJKutf8}  
\def\P{$P$}
\def\B{$B_{\perp}$}

\def\k{$\kappa$}

\def\M{M$_\odot$}
\def\Z{Z$_\odot$}
\def\kms{km\,s$^{-1}$}

\definecolor{darkgreen}{rgb}{0.0,0.55,0.0}

\newcommand{\eg}[0]{$\textnormal{e.g. }$}
\newcommand{\ie}[0]{$\textnormal{i.e. }$}

\graphicspath{{./}{figures/}}

\begin{document}


\title{SN~2018bsz: significant dust formation in a nearby superluminous supernova}

\correspondingauthor{T.-W.~Chen}
\email{twchen@astro.ncu.edu.tw}

\author[0000-0002-1066-6098]{T.-W.~Chen \begin{CJK*}{UTF8}{bsmi}(陳婷琬)\end{CJK*}} 
\affiliation{Graduate Institute of Astronomy, National Central University, 300 Jhongda Road, 32001 Jhongli, Taiwan}
\affiliation{The Oskar Klein Centre, Department of Astronomy, Stockholm University, AlbaNova, SE-10691 Stockholm, Sweden} \affiliation{Max-Planck-Institut f{\"u}r Extraterrestrische Physik, Giessenbachstra\ss e 1, 85748, Garching, Germany}

\author[0000-0003-1325-6235]{S.~J.~Brennan}
\affiliation{The Oskar Klein Centre, Department of Astronomy, Stockholm University, AlbaNova, SE-10691 Stockholm, Sweden} 
\affiliation{School of Physics, O'Brien Centre for Science North, University College Dublin, Belfield, Dublin 4, Ireland}
\author{R.~Wesson}
\affiliation{Department of Physics and Astronomy, University College London, United Kingdom}
\affiliation{School of Physics and Astronomy, Cardiff University, Queen's Building, The Parade, Cardiff, CF24 3AA, UK}
\author[0000-0003-2191-1674]{M.~Fraser}
\affiliation{School of Physics, O'Brien Centre for Science North, University College Dublin, Belfield, Dublin 4, Ireland}
\author{T.~Schweyer}
\affiliation{The Oskar Klein Centre, Department of Astronomy, Stockholm University, AlbaNova, SE-10691 Stockholm, Sweden}
\author[0000-0002-3968-4409]{C.~Inserra}
\affiliation{School of Physics and Astronomy, Cardiff University, Queen's Building, The Parade, Cardiff, CF24 3AA, UK}
\author{S.~Schulze}
\affiliation{Center for Interdisciplinary Exploration and Research in Astrophysics (CIERA), Northwestern University, 1800 Sherman Ave., Evanston, IL 60201, USA} 
\author[0000-0002-2555-3192]{M.~Nicholl} 
\affiliation{Astrophysics Research Centre, School of Mathematics and Physics, Queen's University Belfast, Belfast BT7 1NN, UK}  
\author[0000-0003-0227-3451]{J.~P.~Anderson}
\affil{European Southern Observatory, Alonso de C\'ordova 3107, Casilla 19, Santiago, Chile}
\author[0000-0003-1039-2928]{E. Y.~Hsiao \begin{CJK*}{UTF8}{bsmi}(蕭亦麒)\end{CJK*}}
\affil{Department of Physics, Florida State University, 77 Chieftan Way, Tallahassee, FL 32306, USA}
\author{A.~Jerkstrand}
\affiliation{The Oskar Klein Centre, Department of Astronomy, Stockholm University, AlbaNova, SE-10691 Stockholm, Sweden} 
\author{E.~Kankare}
\affiliation{Tuorla Observatory, Department of Physics and Astronomy, FI-20014 University of Turku, Finland}
\author{E.~C.~Kool}
\affiliation{The Oskar Klein Centre, Department of Astronomy, Stockholm University, AlbaNova, SE-10691 Stockholm, Sweden} 
\author{T.~Kravtsov}
\affiliation{Tuorla Observatory, Department of Physics and Astronomy, FI-20014 University of Turku, Finland}
\affiliation{Finnish Centre for Astronomy with ESO (FINCA), FI-20014 University of Turku, Finland}
\author{H.~Kuncarayakti}
\affiliation{Tuorla Observatory, Department of Physics and Astronomy, FI-20014 University of Turku, Finland}
\affiliation{Finnish Centre for Astronomy with ESO (FINCA), FI-20014 University of Turku, Finland}
\author{G.~Leloudas}
\affiliation{DTU Space, Department of Space Research and Space Technology, Technical University of Denmark, Elektrovej 327, 2800 Kgs. Lyngby, Denmark}
\author[0000-0003-1449-7284]{C.-J.~Li \begin{CJK*}{UTF8}{bsmi}(李傳睿)\end{CJK*}}
\affiliation{Graduate Institute of Applied Physics, National Chengchi University, Taipei 116026, Taiwan}
\affiliation{Institute of Astronomy and Astrophysics, Academia Sinica, No. 1, Sec. 4, Roosevelt Rd., Taipei 106216, Taiwan}
\author[0000-0002-5529-5593]{M.~Matsuura}
\affiliation{School of Physics and Astronomy, Cardiff University, Queen's Building, The Parade, Cardiff, CF24 3AA, UK}
\author[0000-0003-4663-4300]{M.~Pursiainen}
\affiliation{Department of Physics, University of Warwick, Gibbet Hill Road, Coventry CV4 7AL, UK}
\author{R.~Roy}
\affiliation{Institute of Astronomy Space and Earth Science (IASES), P 177, CIT Road, Scheme 7m, Kolkata-700054, West Bengal, India}
\author[0000-0002-4794-6835]{A.~J.~Ruiter}
\affiliation{Mathematical Sciences Institute, Australian National University, Australia}
\author[0000-0002-1214-770X]{P.~Schady$^\dagger$}
\affiliation{Department of Physics, University of Bath, Claverton Down, Bath, BA2 7AY, UK}
\author[0000-0002-5044-2988]{I.~R.~Seitenzahl}
\affiliation{Research School of Astronomy and Astrophysics, Australian National University, Canberra, ACT 2611, Australia}
\affiliation{Mathematical Sciences Institute, Australian National University, Australia}
\author[0000-0003-1546-6615]{J.~Sollerman}
\affiliation{The Oskar Klein Centre, Department of Astronomy, Stockholm University, AlbaNova, SE-10691 Stockholm, Sweden}
\author[0000-0003-3433-1492]{L.~Tartaglia}
\affiliation{INAF - Osservatorio Astronomico d'Abruzzo, via Mentore Maggini snc, I-64100 Teramo, Italy}
\author{L.~Wang}
\affiliation{George P. and Cynthia Woods Mitchell Institute for Fundamental Physics \& Astronomy, Texas A\&M University, Department of Physics and Astronomy, 4242 TAMU, College Station, TX 77843, USA}
\author[0000-0001-9320-4958]{R.~M.~Yates}
\affiliation{Centre for Astrophysics Research, University of Hertfordshire, Hatfield, AL10 9AB, UK}
\author{S.~Yang \begin{CJK*}{UTF8}{gbsn}(杨圣)\end{CJK*}}
\affiliation{Henan Academy of Sciences, Zhengzhou 450046, Henan, China}
\affiliation{The Oskar Klein Centre, Department of Astronomy, Stockholm University, AlbaNova, SE-10691 Stockholm, Sweden}

\author[0000-0003-1637-9679]{D.~Baade}
\affiliation{European Organisation for Astronomical Research in the Southern Hemisphere (ESO), Karl-Schwarzschild-Str.\ 2, 85748 Garching b. M{\"u}nchen, Germany}
\author[0000-0003-1604-2064]{R.~Carini}
\affiliation{INAF - Osservatorio Astronomico di Roma, via Frascati 33, I-00078 Monte Porzio Catone, Italy}
\author[0000-0002-3653-5598]{A.~Gal-Yam}
\affiliation{Benoziyo Center for Astrophysics and the Helen Kimmel Center for Planetary Science, Weizmann Institute of Science, 76100 Rehovot, Israel}
\author{L.~Galbany}
\affiliation{Institute of Space Sciences (ICE, CSIC), Campus UAB, Carrer de Can Magrans, s/n, E-08193 Barcelona, Spain}
\affiliation{Institut d'Estudis Espacials de Catalunya (IEEC), Edifici RDIT, Campus UPC, 08860 Castelldefels (Barcelona), Spain}
\author{S.~Gonz\'alez-Gait\'an}
\affiliation{Instituto de Astrof\'isica e Ci\^encias do Espa\c{c}o, Faculdade de Ci\^encias, Universidade de Lisboa, Ed. C8, Campo Grande, 1749-016 Lisbon, Portugal}
\author[0000-0002-1650-1518]{M.~Gromadzki}
\affiliation{Astronomical Observatory, University of Warsaw, Al. Ujazdowskie 4, 00-478 Warszawa, Poland}
\author[0000-0003-2375-2064]{C.~P.~Guti\'errez}
\affiliation{Institut d'Estudis Espacials de Catalunya (IEEC), Edifici RDIT, Campus UPC, 08860 Castelldefels (Barcelona), Spain}
\affiliation{Institute of Space Sciences (ICE, CSIC), Campus UAB, Carrer de Can Magrans, s/n, E-08193 Barcelona, Spain}
\author{R.~Kotak}
\affiliation{Department of Physics \& Astronomy, University of Turku, Vesilinnantie 5, Turku FI-20014, Finland}
\author[0000-0002-9770-3508]{K.~Maguire}
\affiliation{School of Physics, Trinity College Dublin, The University of Dublin, College Green, Dublin 2, Ireland}
\author{P.~A.~Mazzali}
\affiliation{Astrophysics Research Institute, Liverpool John Moores University, IC2, Liverpool Science Park, 146 Brownlow Hill, Liverpool L3 5RF, UK} \affiliation{Max-Planck-Institut f{\"u}r Astrophysik, Karl-Schwarzschild Str. 1, D-85748 Garching, Germany}
\author[0000-0003-3939-7167]{T.~E.~M{\"u}ller-Bravo}
\affiliation{School of Physics, Trinity College Dublin, The University of Dublin, College Green, Dublin 2, Ireland}
\author[0000-0003-2814-4383]{E.~Paraskeva}
\affiliation{IAASARS, National Observatory of Athens, 15236, Penteli, Greece}
\affiliation{Department of Astrophysics, Astronomy Mechanics, Faculty of Physics, National and Kapodistrian University of Athens, 15784 Athens, Greece}
\author[0000-0002-8041-8559]{P.~J.~Pessi}
\affiliation{Facultad de Ciencias Astron\'omicas y Geof\'isicas (FCAG), Universidad Nacional de La Plata (UNLP), Paseo del bosque S/N, 1900, Argentina}
\affiliation{European Southern Observatory, Alonso de C\'ordova 3107, Casilla 19, Santiago, Chile}
\author{G.~Pignata}
\affiliation{Instituto de Alta Investigaci\'on, Universidad de Tarapac\'aca, Casilla~7D, Arica, Chile}
\author{A.~Rau}
\affiliation{Max-Planck-Institut f{\"u}r Extraterrestrische Physik, Giessenbachstra\ss e 1, 85748, Garching, Germany}
\author[0000-0002-1229-2499]{D.~R.~Young}
\affiliation{Astrophysics Research Centre, School of Mathematics and Physics, Queen's University Belfast, Belfast BT7 1NN, UK}

\renewcommand{\thefootnote}{$\dagger$}
\footnotetext{Deceased.}

\begin{abstract}
We investigate the thermal emission and the extinction from dust associated with the nearby ($z = 0.0267$) superluminous supernova (SLSN) 2018bsz. 
Our dataset combines daily cadence, simultaneous optical and near-infrared coverage up to $\sim$100\,days with late time ($+1.7$\,yr) mid-infrared observations.  
At 230\,days after light curve peak, the SN is not detected in the optical, but shows a surprisingly strong near-infrared excess, with $r-J>3$\,mag and $r-K_{\mathrm{s}}>5$\,mag. 
The temporal evolution of the infrared light curve allows us to assess whether the mid-infrared emission originates from newly formed dust, either in the SN ejecta or in a post-shock circumstellar shell, or instead from an infrared echo produced by pre-existing circumstellar or interstellar dust heated by the SN radiation.
Our radiative-transfer modelling shows that the circumstellar and interstellar echo scenarios cannot simultaneously reproduce the observed infrared luminosity and its temporal evolution, and we therefore disfavour these interpretations. In contrast, a scenario in which new dust forms in the SN ejecta at epochs $\gtrsim 200$\,days provides a self-consistent explanation for the evolution of the SN flux.
We can fit the spectral energy distribution well at $+230$\,d with 
$5\times 10^{-4}\,M_{\odot}$ of carbon dust, increasing over the following several hundred days to $10^{-2}\,M_{\odot}$ by $+535$\,d. 
SN~2018bsz is the first SLSN showing evidence for dust formation within the SN ejecta, and appears to form ten times more dust than normal core-collapse SNe at similar epochs. Together with their preference for low-mass, low-metallicity host galaxies, we suggest that SLSNe may be a significant contributor to dust formation in the early Universe.
\end{abstract}



\section{Introduction}
\label{sec:intro}

Core-collapse supernovae (CCSNe) are considered to be significant sites of cosmic dust production (\citealt{1967AnAp...30.1039C, 1970Natur.226...62H}; for more recent reviews see \citealt{2011A&ARv..19...43G, 2018SSRv..214...63S}). Emission from newly formed dust has been seen in both CCSNe \citep[e.g.][]{1989IAUC.4746....1D, 1989LNP...350..164L, 2004A&A...426..963E, 2009ApJ...704..306K,2015ApJ...800...50M} and in CCSN remnants \citep[e.g.][]{2001A&A...369..589D}.  
Additionally, CO has been detected in the near-infrared (NIR) spectra of approximately twenty Type II (hydrogen-rich) SNe \citep[e.g.][]{1987IAUC.4468....2M, 1988ApJ...331L...9E, Spyromilio:1988hk, Lepp:1990cz, Petuchowski:1989gg, 2017hsn..book.2125M, 2021ApJ...908..232R, 2018ApJ...864L..20R, 2019ApJ...887....4D}, and in a few Type Ic (hydrogen-poor) SNe \citep[e.g.][]{2021ApJ...908..232R} about 100 days after the explosion. This is considered to be an indication of chemistry triggered in the SN ejecta, and can be regarded as a precursor to dust formation \citep[e.g.][]{2003ApJ...598..785N, 2018SSRv..214...63S}. 
The amount of newly-formed dust in nearby CCSNe typically ranges between $10^{-5}$ and $10^{-3}\,M_{\odot}$ \citep[see reviews by][]{2011A&ARv..19...43G,2018SSRv..214...63S}, and is usually estimated from observations taken within three years after explosion, or in some cases even earlier.
Dust masses inferred in SN remnants are generally larger, the average mostly ranges between $10^{-1}$ and 1\,$M_{\odot}$ \citep[e.g.][]{2012ApJ...760...96G, 2017MNRAS.465.3309D, 2017MNRAS.465.4044B, 2019MNRAS.483...70C}.
Notably, \citet{2011Sci...333.1258M} used the \textit{Herschel Space Observatory} to observe the far-infrared (FIR) emission of SN~1987A in 2010 and found that a large dust mass of $\sim0.4-0.7$\,$M_{\odot}$ was formed in this SN.
Recently, \textit{JWST} found self-absorption of the ejecta dust in SN~1987A \citep{2024MNRAS.532.3625M}, confirming that earlier ALMA results of large amount of ejecta dust \citep{2019ApJ...886...51C}. 

Observations of the most distant quasars ($z>6$) show evidence for large masses of dust in galaxies in the early Universe \citep[e.g.][]{2003A&A...406L..55B, 2007ApJ...662..927D, 2025NatAs...9..458M}.  This is a strong argument for SNe being important dust producers in the high-$z$ Universe \citep[e.g.][]{2002MNRAS.337..921H,2003MNRAS.343..427M,2004Natur.431..533M,2007MNRAS.378..973B,2011A&A...528A..14G,2014MNRAS.438.2765C}, since other channels such as asymptotic giant branch (AGB) stars (through stellar outflows) will not have had sufficient time to produce a large quantity of dust at such early epochs \citep[e.g.][]{1998ApJ...501..643D,2006A&A...447..553F,2011ApJ...727...63D}. However, it has been suggested that the AGB stars may be an important contributor to dust formation by $z\sim6-7$ \citep{2009MNRAS.397.1661V, 2011MNRAS.416.1916V}. 
Nonetheless, it is still a challenge to account for the large amount of dust in high-$z$ galaxies \citep[e.g.][]{2001MNRAS.325..726T,2001ApJ...562..480C, 2011ApJ...727...63D}, and additional dust production mechanisms are required \citep[e.g.][]{2009MNRAS.396..918M}, such as grain growth in the interstellar medium (ISM) \citep{2003ApJ...598.1017D, 2019A&A...624L..13L}.

Superluminous SNe (SLSNe) are massive stellar explosions which are 10--100 times brighter than normal CCSNe \citep{2011Natur.474..487Q, 2019ARA&A..57..305G, 2019NatAs...3..697I}. SLSNe in general have higher explosion energies and more massive ejecta than normal CCSNe \citep[e.g.][]{2015MNRAS.452.3869N, 2018SSRv..214...59M}, and are potentially very different sites for molecular and dust formation \citep[cf.][]{2003ApJ...598..785N,2009ApJ...703..642C,2010ApJ...713....1C,2011ApJ...736...45N}. SLSNe are suspected to be related to the first generation of stars in the Universe \citep{2009Natur.462..624G} and have so far been found at redshifts as high as $z\approx2$ \citep{2017MNRAS.470.4241P,2018ApJ...854...37S, 2019MNRAS.487.2215A} or even $z\approx4$ \citep{2012Natur.491..228C,2019ApJS..241...16M,2019ApJS..241...17C}. 

At low redshift Universe ($z<2$), the most common Type I (hydrogen-poor) SLSNe predominantly occur in dwarf galaxies with low stellar mass and low metallicity \citep{2011ApJ...727...15N, 2013ApJ...763L..28C, 2015ApJ...804...90L, 2015MNRAS.449..917L, 2016MNRAS.458...84A, 2016ApJ...830...13P, 2017MNRAS.470.3566C, 2018MNRAS.473.1258S}. 
Moreover, the dwarf galaxies hosting Type I SLSNe are often part of interacting systems \citep{2017A&A...602A...9C, 2020A&A...643A..47O}.
Such host galaxies are expected to be similar to galaxies at high redshift (\eg{}\citealt{Heckman+05,Cardamone+09,Bian+16}). 
As Type I SLSNe may arise from more massive progenitors \citep{2017ApJ...835...13J, 2019MNRAS.484.3451M}, they could potentially produce more dust than normal CCSNe at similar metallicity. However, \citet{Gomez24} placed only upper limits on the dust mass in the Type I SLSN~2017gci, namely $0.83\,M_\odot$ at the 3$\sigma$ level and $0.44\,M_\odot$ at the 1$\sigma$ level. A larger sample of Type I SLSNe will therefore be required to assess their role in the early formation and evolution of cosmic dust. By comparison, \citet{2022MNRAS.513.4057S} found that hydrogen-rich SLSNe appear to form dust masses about ten times larger than those observed in normal core-collapse SNe at similar epochs.

SN~2018bsz was classified as the closest Type I SLSN to date, at a redshift of $z=0.0267$ \citep{2018A&A...620A..67A}. It exhibited several unusual features, including a long plateau prior to maximum light, and strong optical emission lines associated with C\,{\sc ii}. Late-time spectra show hydrogen, possibly indicating late-time interaction with an H-rich shell \citep[e.g.][]{2015ApJ...814..108Y,2017ApJ...848....6Y}. 
\citet{2022A&A...666A..30P} had presented detailed analyses of extensive spectroscopic datasets of the source, and more photometric datasets will be presented elsewhere (Roy et al. in prep.).
While in this paper, we focus on the late-time behaviour and mid-infrared (MIR) properties of SN~2018bsz, in order to investigate what role SLSNe play in cosmic dust formation.
For consistency, we adopt the same basic parameters for SN~2018bsz as used in the early-phase study by \citet{2018A&A...620A..67A}. We therefore assume a Hubble constant $H_{0}=73$\,km\,s$^{-1}$\,Mpc$^{-1}$, $\Omega_{\Lambda}=0.73$ and $\Omega_{\rm M}=0.27$ \citep{2007ApJS..170..377S}. The corresponding luminosity distance $D_{L}=111$\,Mpc and a distance modulus of $\mu=35.23\pm0.02$\,mag are used for SN~2018bsz. The rest-frame phase is with respect to the {\textit r}-band peak epoch (MJD = 58267.5). Throughout this paper all magnitudes are reported in the AB system. 

\section{Observations}
\label{sec:obs}

\subsection{Imaging from ultraviolet to mid-infrared wavelengths}
\label{sec:imaging}

We obtained multi-band light curves for SN~2018bsz from the ultraviolet (UV) to the MIR with several telescopes. We used the UV/optical telescope (UVOT; \citealt{2005SSRv..120...95R}) on board the Neil Gehrels \emph{Swift} Observatory \citep{2004ApJ...611.1005G}, under three target-of-opportunity programs\footnote{P.I.s Brown, Dong, Roy and Schulze}; the Gamma-Ray Burst Optical/Near-Infrared Detector \citep[GROND;][]{2008PASP..120..405G} mounted on the 2.2-m MPG telescope at ESO's La Silla Observatory in Chile, as part of the GREAT survey \citep{2018ApJ...867L..31C}; and the infrared spectrograph and imaging camera Son of ISAAC (SOFI) on the 3.58-m New Technology Telescope (NTT) in the framework of the extended Public ESO Spectroscopic Survey of Transient Objects \citep[ePESSTO;][]{2015A&A...579A..40S} program.
We were also awarded Director's Discretionary Time on the \textit{Spitzer Space Telescope} \citep{2004ApJS..154....1W} using the Infrared Array Camera (IRAC) instrument \citep{2004ApJS..154...10F} for late-time observations\footnote{Program ID: 14273, PI: Chen}. In addition, archival mid-IR observations of the field of SN~2018bsz were taken by the Near Earth Object Wide-field Infrared Survey Explorer \citep[NEOWISE;][]{2011ApJ...731...53M} project using the {\it Wide-field Infrared Survey Explorer} \citep[\textit{WISE};][]{2010AJ....140.1868W}, obtained as part of the NEOWISE Reactivation survey \citep[NEOWISE-R;][]{2014ApJ...792...30M}. 

\begin{figure*}
\centering
\includegraphics[width=16cm,trim={0.5cm 0cm 0.5cm 0.5cm}]{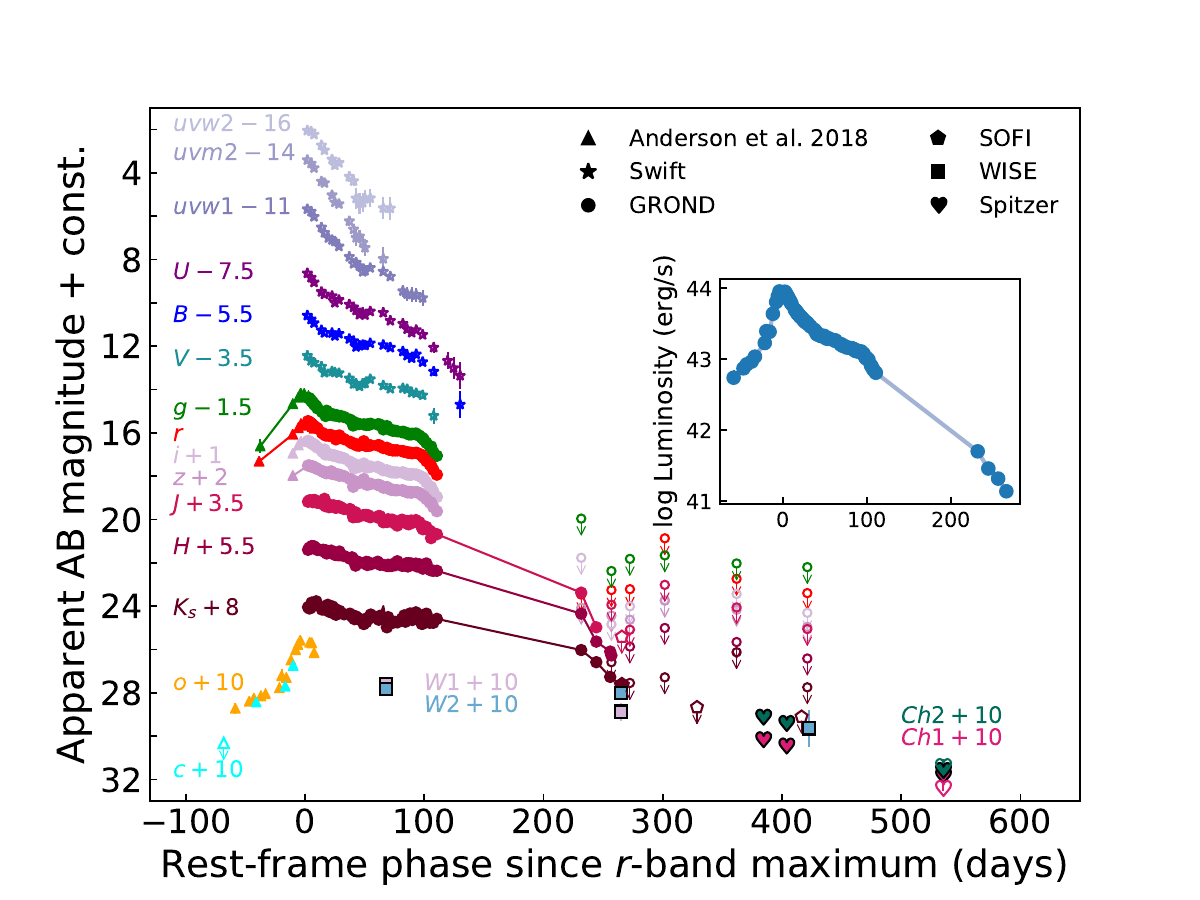}
\caption{Multi-band, host-subtracted light curves of SN~2018bsz. Photometric measurements have not been corrected for Milky Way or host galaxy extinction. 
The pre-peak light curves are taken from \citet{2018A&A...620A..67A}. 
Marker shapes indicate data points taken using different facilities. The open symbols show $3\sigma$ detection limits. The inset panel shows the bolometric light curve of SN~2018bsz (uncertainties are smaller than the point size).
Note that at late-time phases ($>200$\,days), the SN was not detected in the optical but still detected in the NIR bands, with $r-K_{\mathrm{s}}>5$\,mag.
}
\label{fig:phot_colour}
\end{figure*}

We discuss the data reduction, photometric calibration and template subtraction for these images in detail in Appendix\,\ref{sec:app_images}. 
Photometry on our optical and near-infrared (NIR) images was obtained using the custom-built photometry pipeline {\sc AutoPhOT} \citep{2022A&A...667A..62B}; the version of the software used in this work is archived on Zenodo \citep{brennan_2026_21534239}.
All magnitudes have been measured after subtraction of the background host galaxy (Tables\,\ref{tab:phot_swift}, \ref{tab:phot_grond_optical}, \ref{tab:phot_grond_nir}, \ref{tab:phot_sofi}, \ref{tab:phot_spitzer}, \ref{tab:phot_neowise}).
Figure~\ref{fig:phot_colour} shows our assembled multi-band light curves for SN~2018bsz from 2 days after peak ($+2$\,d) to $+535$\,d, combined with the pre-peak ATLAS light curves from \citet{2018A&A...620A..67A}. 

\subsection{Optical and NIR spectroscopy}
\label{sec:spectroscopy}

We also obtained a number of optical and NIR spectra for SN~2018bsz.
At $+17$\,d, we used the Spectrograph for INtegral Field Observations in the Near Infrared (SINFONI) instrument \citep{2003SPIE.4841.1548E}, mounted on the ESO 8-m Very Large Telescope (VLT), to observe SN~2018bsz and the surrounding H\,{\sc ii} regions. 
A spectrum at $+77$\,d was taken with the integral field unit (IFU) Wide-Field Spectrograph (WiFeS) \citep{2007Ap&SS.310..255D}, mounted on the Australian National University (ANU) 2.3-m telescope. 
A $+108$\,d spectrum was observed with the ESO Faint Object Spectrograph and Camera (EFOSC2) \citep{1984Msngr..38....9B} on the NTT through the ePESSTO collaboration.  
After the SN emerged from solar conjunction, we attempted a late-time observation with X-Shooter \citep{2011A&A...536A.105V}, mounted on the VLT, between $+327$\,d and $+383$\,d. No SN trace was detected. Finally we used the Multi-Unit Spectroscopic Explorer \citep[MUSE; ][]{Bacon10} to observe the host-galaxy environment after the SN had faded at $+417$\,d.

\begin{figure*}
    \centering
    \includegraphics[width=18cm]{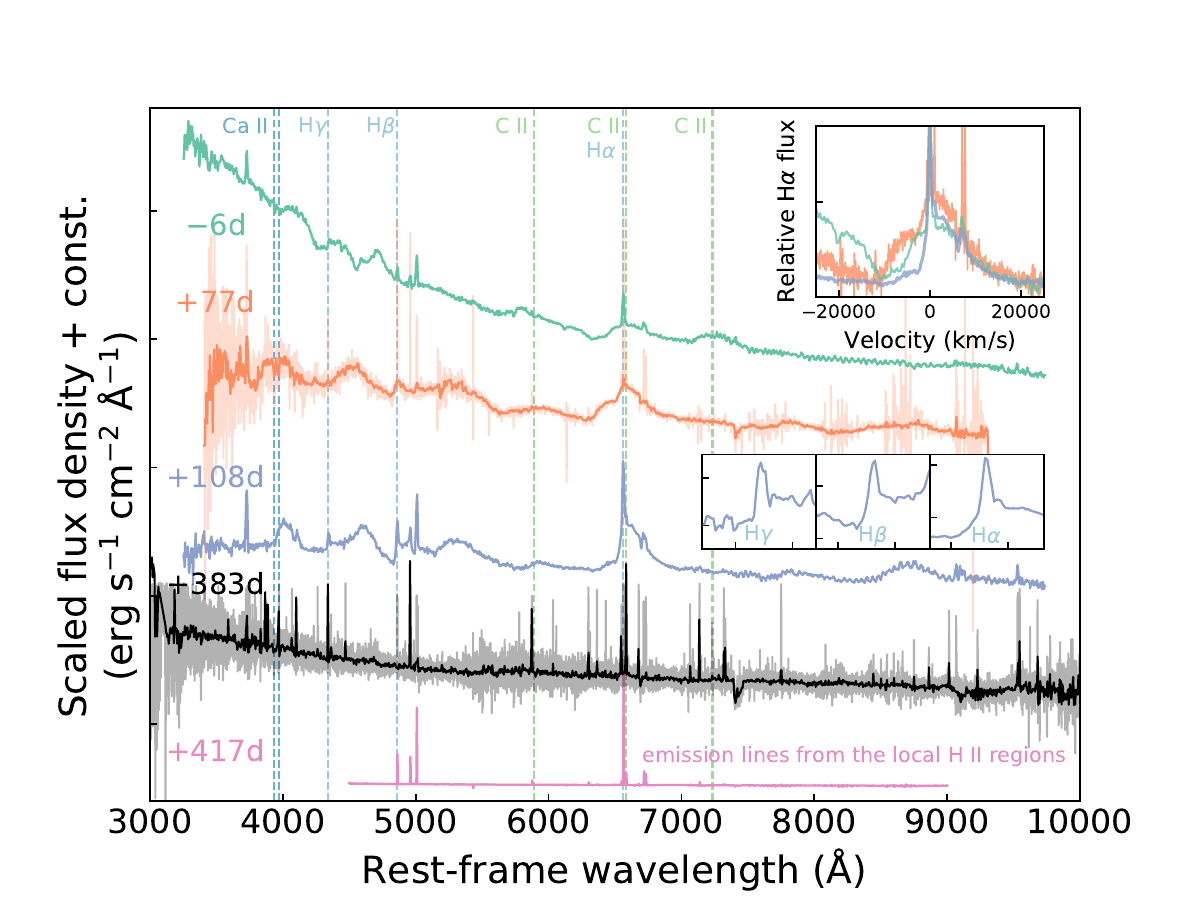}
    \caption{Spectroscopic evolution of SN~2018bsz. The $-6$\,d spectrum is taken from \citet{2018A&A...620A..67A}. The Balmer lines have emerged by $+77$\,d in our WiFeS spectrum (the darker line has been rebinned to 20 \AA). Interestingly, a strong asymmetry in the line profiles of all Balmer lines (see lower inset panels for zoom) appears in the $+108$\,d EFOSC2 spectrum. This spectrum was previously presented by \citet{2022A&A...666A..30P}.
    No SN trace has been detected in the $+383$\,d X-Shooter spectrum (the darker line has been binned to 5 \AA). The $+417$\,d spectrum was extracted from the MUSE cube at the SN position. After subtracting the stellar continuum, we highlight the strong narrow emission lines from the local H\,{\sc ii} regions (pink line). The upper inset shows a zoom in around H$\alpha$, where the flux has been scaled to match the same continuum level as on the red side of the line. The spectra shown in this figure are provided as Data behind the Figure.}
    \label{fig:spec}
\end{figure*}

The instrumental configurations are listed in Table\,\ref{tab:log_spec}, and we describe the data reduction in detail in Appendix\,\ref{sec:app_spectroscopy}.
We show these optical spectra of SN~2018bsz in Fig.\,\ref{fig:spec}, while we plot the NIR spectra in Fig.\,\ref{fig:SINFONI_XS}. The complete set of spectra obtained for SN~2018bsz is presented and analysed in \citet{2022A&A...666A..30P}. 
The spectra shown in Figures~\ref{fig:spec} and~\ref{fig:SINFONI_XS} are publicly available through WISeREP \citep{2012PASP..124..668Y} and are provided as Data behind the Figure.

\section{Observational characteristics of SN~2018bsz}
\label{sec:characteristics}

\subsection{Extinction}
\label{sec:extinction}

Several sources can contribute to the extinction towards SN~2018bsz: a foreground (Milky Way) component; interstellar extinction within the host galaxy 2MASX J16093905-3203443; and (potentially) any circumstellar extinction surrounding the progenitor. The first of these components can be determined from Milky Way dust maps to be $E(B-V){\rm_{MW}}=0.214$\,mag \citep{2011ApJ...737..103S}. The latter two components are more problematic, and we consider a number of indirect approximate methods to measure them.

We extracted the host galaxy spectrum from our MUSE datacube at the position of SN~2018bsz using a circular aperture with 0\arcsec.6 radius.  
After correcting for Milky Way extinction, we fitted and removed the stellar continuum using {\sc starlight} \citep{2005MNRAS.358..363C}, and then measured the Balmer decrement in the spectrum. 
Assuming that the intrinsic value of the Balmer decrement of H$\alpha$/H$\beta$ is 2.86 \citep{1989agna.book.....O}, we found $E(B-V){\rm_{host}}$ to be $\sim0.32$\,mag. The host of SN~2018bsz is part of an interacting galaxy pair, where in principle one may expect relatively high amounts of dust from active star formation \citep[e.g.][]{2013ApJ...768...90L}. 

However, if we were to apply such a high extinction correction to the SN, the spectral energy distribution (SED) fits would become poor assuming a blackbody, and the uncertainty of the derived temperature of SN~2018bsz increases by 25\%. It  would become $>18800$\,K at 3--20 days after the peak. The inferred temperature is higher (by $\sim5,000$\,K) and outside of the $3\sigma$ temperature range of other Type I SLSNe at the same phase (see Fig.\,\ref{fig:BB_temp}; \citealt{2018MNRAS.475.1046I}).   

We tested a range of different extinctions for SN~2018bsz and found that $E(B-V){\rm_{host}}$ between 0.03 and 0.10\,mag can give a reasonable agreement with the temperature evolution of other SLSNe at 3--20 days after the peak.
\cite{2018A&A...620A..67A} used $E(B-V){\rm_{host}}=0.04$\,mag derived from the equivalent width of the sodium doublet (Na I D) lines \citep{2012MNRAS.426.1465P}. For consistency, we also adopt this value here, corresponding to host $A_V=0.13$\,mag assuming the Galactic $R_V=3.1$ \citep{1989ApJ...345..245C}. In addition, we also obtained a similar $A_V=0.15$\,mag from our {\sc starlight} best fit results. Moreover, we used the spectrum of SN~2018bsz \citep{2018A&A...620A..67A} taken before our first epoch of multi-band photometry to fit a blackbody. At phase $-8$\,d, adopting host $A_V=0.13$\,mag, the inferred temperature is $\sim13700$\,K; while assuming host $A_V=1$\,mag, it becomes $\sim41000$\,K which is not physically possible.

SN~2018bsz seems to lie towards the outskirts of its host and could well be positioned in the foreground, which may explain the relatively moderate reddening ($A_V=0.13$\,mag) compared to that inferred from the Balmer decrement ($A_V\sim 1$\,mag). 
Since the Balmer-decrement measurement is based on an aperture covering multiple H\,{\sc ii} regions, it does not necessarily reflect the line‐of‐sight extinction at the SN~2018bsz site. 
We therefore adopt $A_V=0.13$\,mag for the host-galaxy extinction, and $A_V=0.66$\,mag for the Milky Way extinction when correcting the SN photometry and spectroscopy, unless otherwise noted.

\subsection{Light curve and bolometric light curve}
\label{sec:lightcurves}

We have observational coverage for SN~2018bsz from UV, optical, NIR, and (for the first time for an SLSN) MIR wavelengths. Figure\,\ref{fig:phot_colour} shows our multi-band light curve for SN~2018bsz. 
The daily cadence over more than 100 days makes SN~2018bsz the best monitored SLSN to date and it is also the first SLSN with daily contemporaneous 7-band coverage from optical to NIR, thanks to GROND (Fig.\,\ref{fig:grond_phot_roomin}).

The early time evolution of SN~2018bsz was presented in \cite{2018A&A...620A..67A}, showing an ATLAS $o$-band light curve with a long plateau slowly rising for $\sim43$ days during the pre-maximum phase. 
The SN then displayed a steeper rise in magnitude for the next $\sim18$ days in all optical bands. 
It reached a peak brightness at an absolute magnitude of $-20.3$\,mag in the $r$ band. 
In this paper, we focus on the evolution after the peak, where the $r$-band light curve displays three phases. 
Initially, the SN's brightness fell steeply for 18 days, after which it entered a long, slowly diminishing plateau lasting 78 days. Finally, in the 15 days prior to solar conjunction, it dimmed by 1\,mag. The evolution was similar in the other optical bands, but with a shallower decline in the NIR. After the SN emerged from behind the Sun it was no longer detected at optical wavelengths, with $r>23.5$\,mag on $+232$\,d. At the same time, surprisingly, we found the SN was still detected in the NIR bands with $J\sim19.9$, $H\sim18.9$ and it was brightest in $K_{\mathrm{s}}$ with $\sim18.0$\,mag. 

We constructed a bolometric light curve of SN~2018bsz (see the inset panel in Fig.\,\ref{fig:phot_colour}) integrating over the full wavelength range from the UV to MIR. 
We have relatively sparse MIR coverage, hence we smoothly interpolate the MIR data between the epochs $+68$\,d and $+265$\,d. Although there is no MIR data before 68 days, the fractional MIR luminosity at $+68$\,d (only 1\%) suggests that the early time contribution of the MIR is negligible.  
We describe in more detail the method used for constructing the bolometric light curve of SN~2018bsz and estimating the relative fractions of energy emitted in the different wavelength ranges in Appendix\,\ref{sec:app_bol_cons}.

After peak luminosity ($\sim 9\times 10^{43}\,\mathrm{erg\,s^{-1}}$), the bolometric light curve declines rapidly for 16 days, by which time the luminosity has decreased by 
$\sim50$\% (to $\sim 4.7\times 10^{43}\,\mathrm{erg\,s^{-1}}$). 
The bolometric light curve then continues to decline slowly; between $+17$\,d and $+40$\,d, it drops another $\sim50$\% in luminosity. Subsequently, from $+41$\,d to $+96$\,d, SN~2018bsz further slows its decline rate, taking around twice as long to fade by another $\sim 50$\%.
At $+98$\,d, the light curve shows a sudden drop in luminosity of roughly 40\% in less than 13 days (to $\sim 6.5\times 10^{42}\,\mathrm{erg\,s^{-1}}$).
Unfortunately our observing campaign was terminated at this epoch as SN~2018bsz went behind the Sun.
When the SN became observable again at $+232$\,d, it was 13 times fainter ($\sim 5\times 10^{41}\,\mathrm{erg\,s^{-1}}$) than the last observation at $+111$\,d. At late-time phases, there is no measurable contribution from the UV and optical to the luminosity, which is instead dominated by the NIR and MIR emission.

\begin{figure*}
\centering
\includegraphics[width=\textwidth]{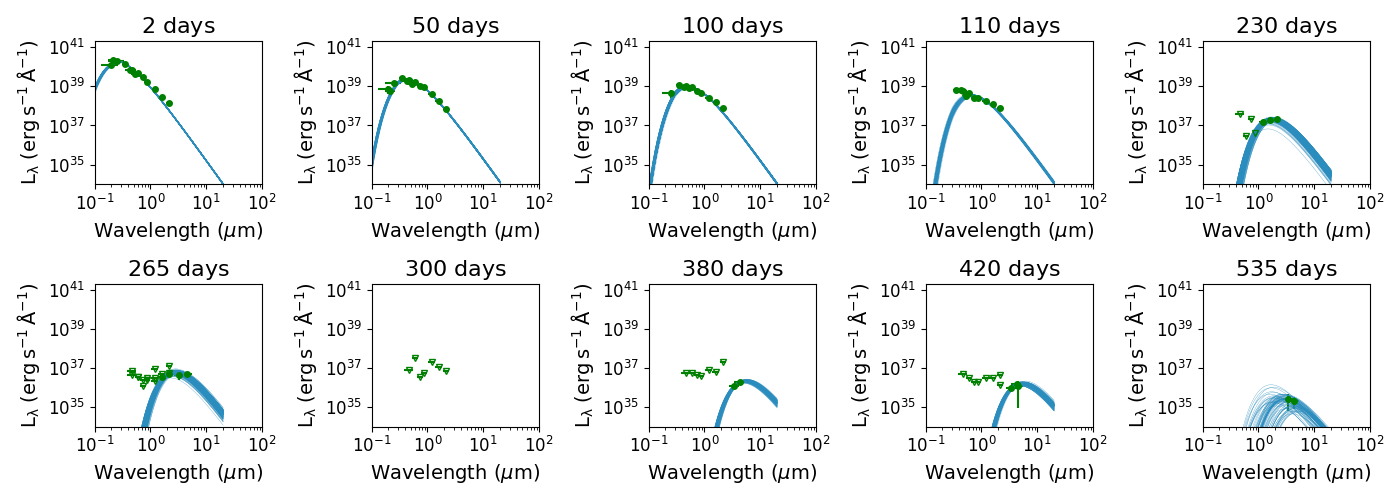}
\caption{SED fitting for SN~2018bsz at selected epochs.
The luminosities were converted from the observed magnitudes after correcting for the Milky Way extinction (green circles) while empty triangles denote 3$\sigma$ upper limits. The corresponding blackbody temperatures and radii are listed in Table\,\ref{tab:bbody}.
}
\label{fig:sed_fitting}
\end{figure*}

To construct representative spectral energy distributions (SEDs) for multi-epoch fitting in Fig.\,\ref{fig:sed_fitting}, we selected a series of photometric epochs that sample the key phases of the light-curve evolution and combine data from UV to IR wavelengths (Tables\,\ref{tab:phot_swift}, \ref{tab:phot_grond_optical}, \ref{tab:phot_grond_nir}, \ref{tab:phot_sofi}, \ref{tab:phot_spitzer}, \ref{tab:phot_neowise}).
The early photometric epochs were chosen to represent key stages of the light-curve evolution and to ensure sufficient wavelength coverage for the SED fitting. Specifically, we selected the first GROND epoch at $+2$\,d to characterize the early post-maximum phase, the $+50$\,d and $+100$\,d epochs to sample the middle and the end of the plateau phase, and another epoch around $+110$\,d corresponding to the last GROND observation obtained before the solar conjunction, when the light curve was declining. The \textit{Swift}/UVOT magnitudes closest in time to each of these GROND epochs were included to extend the SED coverage into the UV bands.

For the late-time evolution, we focus on the NIR and MIR observations. The $+230$\,d epoch corresponds to the first GROND observation after the solar conjunction. The $+265$\,d epoch combines SOFI and NEOWISE data, together with GROND photometry taken near $+257$\,d and $+273$\,d. The $+300$\,d epoch is represented by GROND upper limits, while the $+380$\,d epoch includes \textit{Spitzer} magnitudes and GROND limits obtained at $+362$\,d. At $+420$\,d, \textit{Spitzer} and NEOWISE detections are supplemented by SOFI and GROND upper limits between $+417$\,d and $+423$\,d. Finally, the $+535$\,d epoch includes the \textit{Spitzer} photometry, providing the latest constraints on the infrared emission. We also use these selected epochs for the dust modelling in Sec.\,\ref{sec:model}. 

We investigated if there is a two-component SED by fitting the photometry at the early epochs with a blackbody function. We found that we do not require two distinct components at those epochs, and instead a single blackbody can fit the observed SED well (see Fig.\,\ref{fig:sed_fitting}). Furthermore, we fitted the SED with a blackbody using a Monte Carlo method, and listed the inferred blackbody temperature and radius evolution in Table\,\ref{tab:bbody}.

\subsection{Late-time Hydrogen from CSM}
\label{sect:Hspec}

The spectroscopic evolution of SN~2018bsz based on spectra obtained at key phases is presented in Fig.\,\ref{fig:spec}.
At early phases, \cite{2018A&A...620A..67A} suggested that no hydrogen was visible in the spectra of SN~2018bsz, and that the broad, strong feature at $\sim$6500 \AA\ was not H$\alpha$ but rather C{\sc ii}  $\lambda6580$ at a velocity of $-11000$\,km\,s$^{-1}$. \citeauthor{2018A&A...620A..67A} suggested that the absence of H$\beta$ further strengthened the claim of SN~2018bz being hydrogen-poor. The presence of other C{\sc ii} features at 5890, 7234\,\AA\ meant that carbon was a plausible identification for these lines. In our $+17$\,d NIR SINFONI spectrum (Fig.\,\ref{fig:SINFONI_XS}), we see no broad SN emission at the wavelength of Pa$\alpha$, confirming the lack of spectral signatures of H in the ejecta around maximum light. 

By $+77$\,d, however, it is clear that hydrogen is indeed present in the spectra of SN~2018bsz. 
At $+108$\,d, a peculiar line profile characterised by a sharp cut off in flux in the blue is seen in H$\alpha$, and an identical profile is also seen in H$\beta$ and H$\gamma$, 
making this a secure identification. We interpret the late-time appearance of hydrogen as a sign of interaction between the ejecta and H-rich circumstellar material (CSM). 
Somewhat unusually, by $+108$\,d the Balmer line profiles are strongly asymmetric, with the blue side of the line being suppressed. 
The asymmetric shape of the line profiles can be explained by asymmetric shape of the CSM where these lines are emitted, most likely from a disk or torus \citep{2022A&A...666A..30P}. 
Moreover, \citet{2021MNRAS.503..312M} detect strong polarization in SN~2018bsz during the post‐peak phase, providing independent evidence for asymmetric geometry.

\subsection{Colour evolution and NIR excess}
\label{sec:colours}

During the first 75 days after maximum light, 
the colours of SN~2018bsz evolve slowly towards the red in most bands.
If the steep optical decline around +100~d was caused by extinction from such dust, then we might have expected substantial reddening to be observed in the \textit{(g$-$r)}, \textit{(g$-$i)} and \textit{(g$-$z)} colour evolution between $+98$\,d and $+111$\,d. However, we see no strong colour changes as a function of wavelength during this period, with \textit{(g$-$r)}$\sim0.4\Rightarrow0.4$, \textit{(g$-$i)}$\sim0.2\Rightarrow0.2$ and \textit{(g$-$z)}$\sim0.3\Rightarrow0.4$.  
Similarly, the appearance of asymmetric Balmer profiles (Sec. \ref{sect:Hspec}) does not coincide with any strong change in the colours of SN~2018bsz. Therefore, the suppression of the blue emission is probably not due to dust extinction.

At $+108$\,d, the spectrum of SN~2018bsz appears to be becoming optically thin (e.g. we see the disappearance of the absorption component in Ca).
We therefore consider that the contemporaneous luminosity drop in all optical bands is most likely due to the ejecta transitioning to the nebular phase. Accompanying this, we see a slight redward colour evolution, similar to what is seen in other SNe as they become nebular (for example in SN~2004et; \citealt{2010MNRAS.404..981M, 2023MNRAS.523.6048S}). As for the earlier phases, we do not require any dust component to explain this evolution. 

After the SN emerged from solar conjunction, it was not detected in deep imaging in the optical bands, but was still visible in the NIR bands. At $+232$\,d, the colours were $r-J>3.1$, $r-H>4.1$ and $r-K_{\mathrm{s}}>4.9$\,mag (the magnitudes are corrected for the Milky Way and the host galaxy dust extinction). 
Moreover, the SN shows a prominent $K_{\mathrm{s}}$-band excess (see Fig.\,\ref{fig:cutouts}). 
We measured a colour $J-K_{\mathrm{s}}=1.73\pm0.21$\,mag at $+232$\,d, and an even redder colour $J-K_{\mathrm{s}}=2.77\pm0.30$\,mag at $+244$\,d. The SOFI photometry at $+266$\,d (with a colour $J-K_{\mathrm{s}}>2.15$\,mag) confirms this $K_{\mathrm{s}}$-band excess from GROND.

This rapid evolution to red colours suggests either that new dust is condensing in the SN ejecta, or newly formed dust in the circumstellar envelope, or that there is an IR light echo from pre-existing circumstellar dust or ISM heated by the SN. We investigate those scenarios in detail in Sec.\,\ref{sec:model}.

\begin{deluxetable}{cccccc}
\tablecaption{Blackbody temperature and radius evolution at selected epochs.
\label{tab:bbody}}
\tablewidth{0pt}
\tabletypesize{\scriptsize}
\tablehead{
\colhead{Phase} & 
\colhead{Temperature} &
\colhead{Radius} & \\
\colhead{(day)} &
\colhead{(K)} &
\colhead{($10^{15}$cm)} &
}
\startdata
$+2.6$ & 12540 ($^{+660}_{-680}$) & 2.01 ($^{+0.18}_{-0.17}$) \\
$+50$  & 6800 ($^{+540}_{-430}$) & 3.07($^{+0.28}_{-0.37}$) \\
$+100$ & 5770 ($^{+350}_{-480}$) & 3.08 ($^{+0.45}_{-0.34}$) \\
$+110$ & 4340 ($^{+350}_{-500}$) & 3.77 ($^{+0.72}_{-0.55}$) \\
$+230$ & 1630 ($^{+290}_{-330}$) & 10.99 ($^{+6.51}_{-5.28}$) \\
$+265$ & 1040 ($^{+220}_{-160}$) & 18.16 ($^{+9.80}_{-12.92}$) \\
$+380$ & 530 ($^{+60}_{-60}$) & 57.09 ($^{+24.71}_{-17.29}$) \\
$+420$ & 530 ($^{+70}_{-60}$) & 48.53 ($^{+22.55}_{-16.12}$) \\
$+535$ & 960 ($^{+480}_{-400}$) & 4.44 ($^{+4.80}_{-3.30}$) \\
\enddata
\end{deluxetable}

\section{Dust modelling results}
\label{sec:model}

Infrared emission from CCSNe may arise from new dust forming in the ejecta, a thermal echo from CSM, a thermal echo from nearby ISM, or newly formed dust in the circumstellar envelope. The origin of the emission (which could be from more than one of these mechanisms) may be inferred from the temporal evolution of the emission. New dust will form at a temperature close to the dust sublimation temperature, and if it is heated by the decay of radioactive elements in the ejecta, its luminosity will fade exponentially. In the case of a SLSN powered by the spin down of a newly formed magnetar \citep{2010ApJ...717..245K}, the energy deposited in the expanding remnant would similarly be expected to decline quickly. On the other hand, SN-heated circumstellar or interstellar dust will in general have a temperature lower than the sublimation temperature of the constituent dust, as it was initially colder, but could be heated to any temperature below the sublimation temperature. The temporal evolution of a light echo is determined by the geometry of the echoing material. Extensive CSM may give rise to very long-lasting echoes, such as those seen to continue many years after the explosions of SN~2002hh \citep{2005ApJ...627L.113B,2006ApJ...649..332M} and SN~1980K \citep{2012ApJ...749..170S}.
Finally, when SN shocks propagate into a circumstellar shell, a cool dense shell (CDS) may form between the forward and reverse shocks. Dust formation in a CDS has been inferred in a number of supernovae, including SN~1998S \citep{2004MNRAS.352..457P}, SN~2005ip\citep{2017MNRAS.466.3021S} and SN~2016iog \citep{2025arXiv251112929P}. Because CDS dust formation results from interaction between supernova ejecta and nearby dense circumstellar material, it is generally expected to begin at much earlier epochs than the several hundred days typical of condensation in expanding ejecta without interaction \citep[e.g.][]{2009MNRAS.392..894A}.

In the following we expand on each of these scenarios in more detail.

\subsection{Interstellar echoes}
\label{sec_ism}

\begin{figure*}
\includegraphics[width=\textwidth]{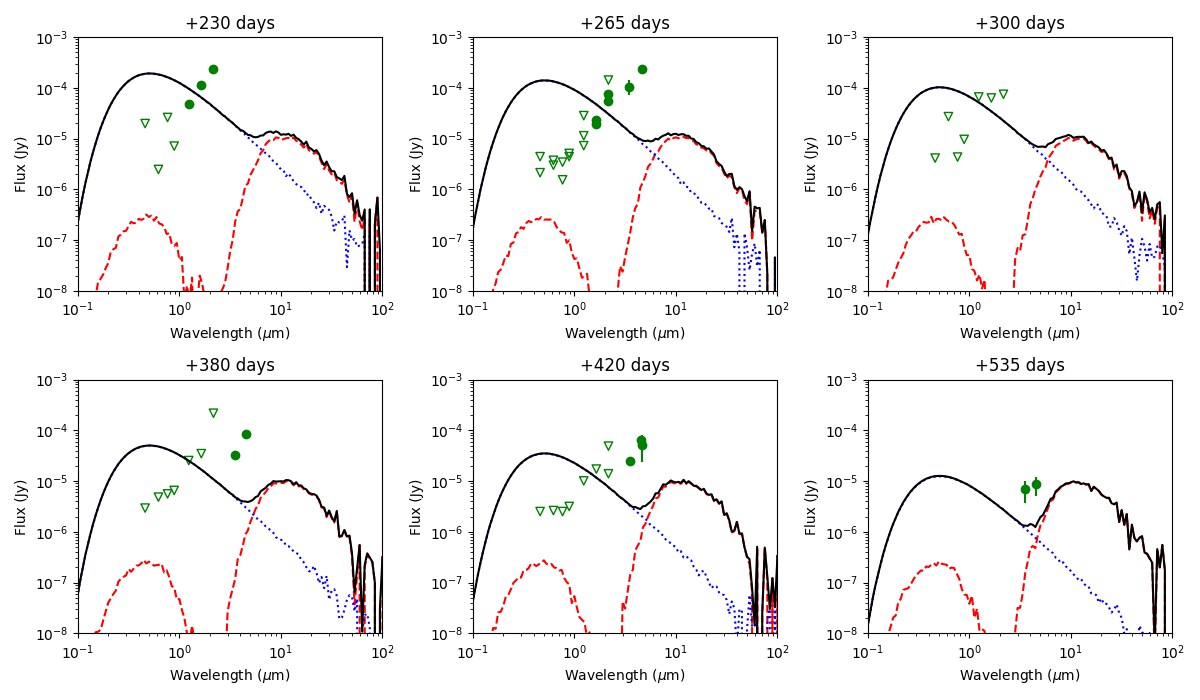}
\caption{The temporal evolution of the emission from a sheet of ISM dust. The predicted SED consists of optical scattered light and the IR echo from dust illuminated by the SN light (red dashed line) and the undelayed emission from the SN (blue dotted line). The IR flux from this assumed ISM is orders of magnitude below the observations, and its flux is quite constant over time. 
The symbols are the same as that used in Fig.\,\ref{fig:sed_fitting}.
The assumed dust model is described in Sec.\,\ref{sec_ism}.}
\label{fig_ISMSEDevolution}
\end{figure*}

We model the evolution of the SED of SN~2018bsz using the radiative transfer code {\sc mocassin} \citep{2003MNRAS.340.1136E, 2005MNRAS.362.1038E}\footnote{We used version 2.02.73 of the code.} and present the modelling results in Fig.\,\ref{fig_ISMSEDevolution}.
We calculated the predicted emission from interstellar and circumstellar echoes using {\sc mocassin}'s option to sum emission from within ellipsoidal regions defined by the light-travel time (see \citealt{2010MNRAS.403..474W} for a description).  We assume that the echoing medium has a low optical depth, so that photons are subject to a single scattering only (multiple scatterings would result in a more complex relationship between the epoch and the geometry of the echoing material). 
The echo SED consists of scattered light and thermal reemission from the echoing material of the SN light at earlier phases, to which we add the undelayed emission from the fading SN ejecta at the epoch of the observation, reddened by its passage through the assumed CSM. We represent the input SN luminosity using the integrated bolometric light curve over the 60-day period in which $L>0.25\times L_{\rm peak}$. This gives a total energy of $2.6\times 10^{50}\,\mathrm{erg}$, and a luminosity of $5\times 10^{43}\,\mathrm{erg\,s^{-1}}$. We assume a blackbody temperature for this input luminosity of $10^{4}$\,K, consistent with the estimated temperature of the SN around the time of its peak luminosity (see Fig.~\ref{fig:BB_temp}).

Figure\,\ref{fig_ISMSEDevolution} shows the temporal evolution of the emission from a volume of dust measuring 10\,pc $\times$ 10\,pc $\times$ 1\,pc, 1\,pc in front of SN~2018bsz, containing $1\,M_\odot$ of amorphous carbon dust. These parameters are arbitrarily chosen, but illustrate the general problems that any ISM echo model would have in matching the observed fluxes and temporal evolution of the SED of SN~2018bsz.

The IR emission from this dust volume is several orders of magnitude below the observed fluxes at early times, and its flux is almost constant with time, while the optical depth of this dust is not high enough to significantly reduce the undelayed optical emission from the SN. Thus the optical limits at early epochs are severely violated while the IR emission is insufficient. The assumed amorphous carbon dust provides greater optical extinction than silicate dust, and so changing the dust composition would only worsen this disagreement with the observations.

Placing the dust closer to the SN does not increase the echo fluxes significantly, as the echoing region then becomes smaller. One could increase the IR flux by invoking some combination of a higher flash luminosity and higher dust mass. However, the dust density required to obscure the SN completely at the observed epochs would be implausible for any interstellar dust cloud or sheet. An ISM echo is thus unlikely to be a significant contributor to the evolving IR emission, except possibly at much later epochs than considered here.

\begin{figure*}
\includegraphics[width=\textwidth,trim={0.5cm 0cm 0.5cm 0.5cm}]{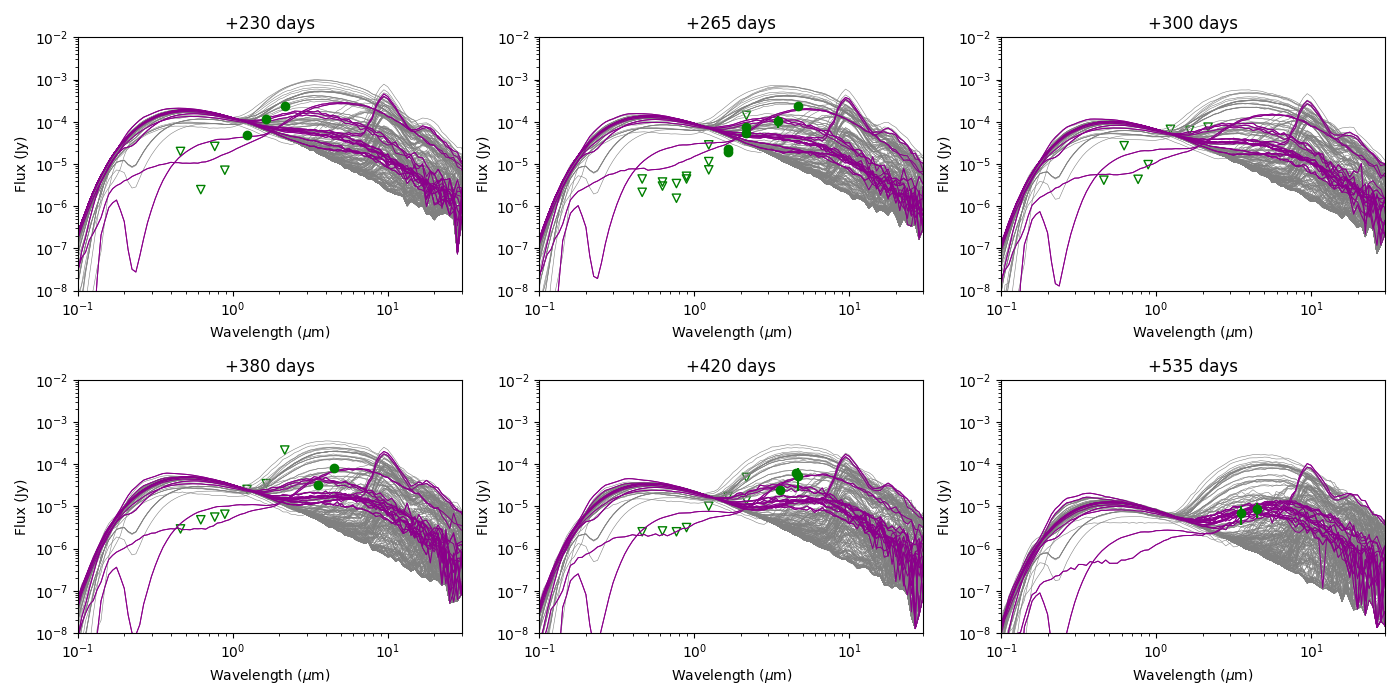}
\caption{Echo SED calculated from a grid of circumstellar shells. 360 models are shown on each panel, covering 40 different shell geometries, three grain sizes and two compositions (silicate and carbon dust, the different is particularly evident in the silicate bump at 10 micron)}. Models which reproduce the NIR/MIR fluxes at $+535$\,d are plotted in magenta; other models are plotted in grey. It can be seen that no model satisfies the observed optical 3$\sigma$ upper limits at early epochs. 
The best-fitting model from this grid is shown in Fig.\,\ref{fig_CSMSEDevolution_bestfit}, which continues this figure by presenting the predicted SED evolution.
The symbols are the same as that used in Fig.\,\ref{fig:sed_fitting}.
\label{fig_CSMSEDevolution_all}
\end{figure*}

\begin{figure*}
\includegraphics[width=\textwidth]{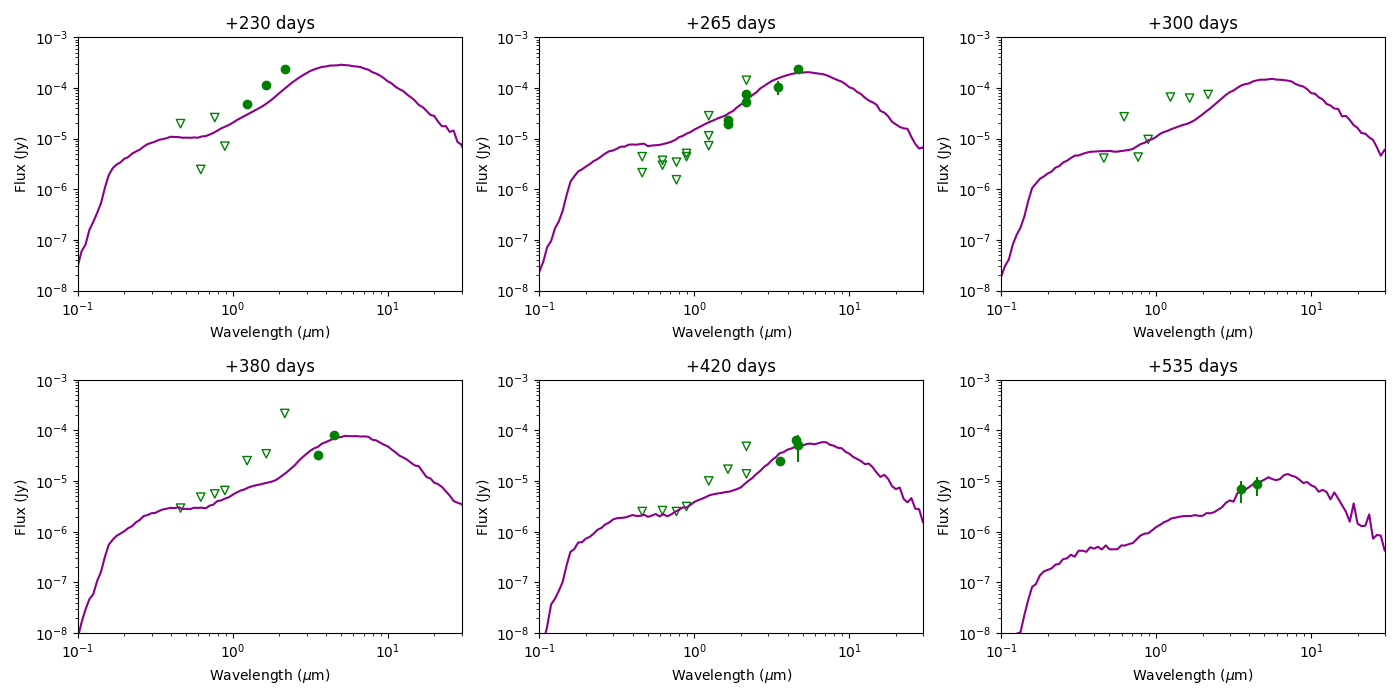}
\caption{Continuation of Fig.\,\ref{fig_CSMSEDevolution_all}, showing only the best-fitting model from that grid. 
Predicted SED evolution for a smooth CSM shell (0.005–0.25\,pc) composed of amorphous carbon grains with $a$=0.1\,$\mu$m. The symbols are the same as in Fig.\,\ref{fig:sed_fitting}. This model yields the best overall agreement with the observed NIR/MIR fluxes among the CSM-shell geometries investigated, but it is disfavoured because it implies an optical depth that would violate the early-time optical $3\sigma$ limits and a inner radius that would lead to ejecta interaction after $\sim$1 yr (Sec. \ref{sec_csm}).}
\label{fig_CSMSEDevolution_bestfit}
\end{figure*}

\subsection{Circumstellar echoes}
\label{sec_csm}

Dense CSM close to the SN may give rise to a significant thermal echo. We constructed models of CSM echoes by assuming a smooth shell in which the density is proportional to $r^{-2}$, with the echo calculated using the same input SN luminosity parameters as the ISM models. The temporal evolution of the echo is determined by the inner and outer radii of the CSM, with the echo ending once the epoch exceeds twice the light-travel time to the outer radius. We calculated the evolution of a light echo from shells of both carbon and silicate dust, with grain radii of 0.01, 0.1 and 1.0\,$\mu$m. We calculated a grid of models for each grain size, consisting of shells with inner radii between 0.001 and 0.1\,pc and outer radii between 2$\times$ and 500$\times$ the inner radius, and we considered total dust masses in the CSM shell of $10^{-3}$ and $10^{-2}\,M_\odot$. If the gas-to-dust ratio is assumed to be 100, this corresponds to a total CSM mass of 0.1 or $1.0\,M_\odot$ respectively. We are unable to find a shell geometry that can account for the evolution of the SED (see Fig.\,\ref{fig_CSMSEDevolution_all}) for three reasons:

Firstly, the temporal evolution is not reproduced. For there to be a CSM echo lasting until $+535$\,d, the shell needs to be very extensive, with at least some material $\sim$270 light days (0.23\,pc) behind the SN. This gives rise to a slowly-declining echo, from which the fluxes at $+230$\,d and $+265$\,d, for example, would not be significantly different. The observations show that over these 35 days, the IR fluxes in fact decline significantly.

Secondly, the temperature evolution of the dust is inconsistent with the observations: newly-heated dust is too cool to reproduce the SED at the earlier epochs, and too warm at the final epoch.

And thirdly, the optical and IR non-detections at $+230$\,d and $+265$\,d cannot be reproduced in most of these echo scenarios; models that predict IR fluxes comparable to those observed also generally predict detectable optical signals from scattered light alone, even before including the undelayed emission from the SN ejecta at the epoch of the observation. Only models with small grains and high dust masses can reproduce the optical non-detections; none of the models which satisfy this constraint give enough IR emission to match the observed SED.

Figure\,\ref{fig_CSMSEDevolution_all} shows the predicted SED fluxes for the subset of echo models which can reproduce the IR observations at $+535$\,d. The SED is the sum of the echo - both scattered optical and reprocessed IR emission - and the undelayed SN emission at the relevant epoch, reddened by the dust shell which is echoing. It can be seen from this figure that echo geometries which give rise to significant emission at $+535$\,d invariably violate the deep optical limits at $+230$\,d and $+265$\,d, in most cases by over an order of magnitude. On the other hand, models which do not violate the optical limits at the earlier epochs do not reproduce the infrared emission at the final epoch.

Figure\,\ref{fig_CSMSEDevolution_bestfit} 
shows the predicted SED for the CSM geometry which provides the closest fit to the observations. The CSM in this model extends from 0.005 to 0.25\,pc, with $10^{-2}\,M_\odot$ of 0.01\,$\mu$m amorphous carbon grains. While the inner edge of the CSM is close to the SN, the maximum dust temperatures we find from this model are $\sim$~1000\,K, too low for the dust to be destroyed. This model gives infrared fluxes at $+380$\,d and later which match the observations reasonably well, but predicts optical emission at $+230$\,d and $+265$\,d which exceeds the observed upper limits by factors of 2--8. Furthermore, this CSM has an optical depth at 0.55\,$\mu$m of 4.3, corresponding to A$_V$=4.67\,mag, where the observations support a value of $\sim$0.1\,mag (Section~\ref{sec:extinction}). One final consideration which rules out this model is that supernova ejecta moving at 5,000\,km\,s$^{-1}$ would reach dust at 0.005\,pc after $\sim$1\,yr, at which point a major rebrightening at all wavelengths would be expected. This is contrary to the observed continuing decline in the bolometric luminosity out to day 535. We therefore conclude that a CSM echo alone cannot reproduce the observed optical-IR evolution of the SED.

Due to the insufficient observational constraints of geometry at late-time phases ($>200$\,days), we have not considered modelling echoes from an asymmetrically-distributed CSM. An asymmetric CSM could give rise to thermal echoes with a temporal evolution more consistent with the rapid fading of the IR emission. However, it remains difficult to see how the optical and IR non-detections at $+230$\,d and $+265$\,d could be reproduced in this scenario, given the contradictory requirements that the amount of dust along the line of sight must be low enough to match the early colours, yet high enough to obscure the fading SN by $+230$\,d.

\subsection{Newly formed dust in circumstellar material}
While the narrow lines in the spectra of SN~2018bsz indicate the presence of CSM, we do not have a robust estimate of the CSM density or radial profile, as these quantities are generally difficult to infer even with detailed modelling. However, typical CSM densities in interacting SNe are on the order of $10^9$--$10^{10}\,\mathrm{cm^{-3}}$.
Some Type IIn SNe (SNe with narrow H line emission), show NIR excesses as early as a few tens of days after explosion \citep[e.g.][]{2004MNRAS.352..457P, Smith:2008dz, Fox:2009cl, 2014Natur.511..326G}.
This early NIR emission is often attributed to newly formed dust in a dense circumstellar shell: while SN shocks propagate into the circumstellar shell, there is a cool region between the forward and reverse shocks where dust can form
\citealp[(e.g.][]{Fox:2010gi}; although see problems with this scenario raised by \citealp{Sarangi.2018}).
The main argument for dust formation in these cases is that the blackbody radius estimated from the infrared excess is much larger than the extent of the ejecta \citep{2014Natur.511..326G}.

In the case of SN\,2018bsz,the blackbody radius at day 265 (corresponding to the epoch with MIR observations) is $1.8\times 10^{16}\,\mathrm{cm}$ (Table 8). This is already a factor three further out than the maximum extent of the CSM ($6.5\times 10^{15}\,\mathrm{cm}$) suggested by \citet{2022A&A...666A..30P}, and in fact is consistent with the radius of the ejecta, if we take it to have a velocity of 8000~\kms. Hence even if dust might have formed in a `circumstellar shell', it will have been largely engulfed by the ejecta, such that it is no longer meaningful. Alternatively, dust in the CSM could be swept up by the ejecta, however to assess how plausible this is would require detailed hydrodynamical modeling that is beyond the scope of this work.

We caution however, that the velocity of the ejecta in SN~2018bsz is highly uncertain - at early times the velocities we observe are associated with CSM rather than ejecta, while SNe also display stratified ejecta where different elements trace different velocity layers \citep{Gutierrez17}. The blackbody radii are similarly uncertain, and as can be see from Fig. \ref{fig:Velocity}, the fits around 200-400 days are based on a small number of IR detections and (mostly) upper limits.

\subsection{New dust forming in the SN ejecta}
\label{sec_new_dust_ejecta}

Given that an ISM echo cannot contribute significantly to the IR fluxes (Fig.\,\ref{fig_ISMSEDevolution}), and an echo from a symmetric CSM is not compatible with the evolution of the SED (Fig.\,\ref{fig_CSMSEDevolution_bestfit}), we consider models of newly-forming dust in the ejecta. We adopt a configuration in which the heating source is distributed throughout an expanding dusty shell, in which the dust density is proportional to $r^{-2}$. The density profile may indeed be steeper than this, but it is computationally difficult to model as steeper exponents require higher grid resolution to sensibly represent. An $r^{-2}$ distribution has also been assumed in previous works carrying out RT modeling of SN ejecta \citep{2007MNRAS.375..753E, 2020ApJ...894..111B}. The temperature of the heating source is not strongly constrained by the observations, given the optical non-detections; we adopt a value of 5,000\,K, based on ejecta temperatures of 3,000--7,000\,K estimated at similar epochs in SN~1987A (\citealt{1993ApJS...88..477W, 2007MNRAS.375..753E}). We assumed a velocity of 1000\,km\,s$^{-1}$ at the inner edge of the ejecta, and an outer radius 5 times the inner radius. The predicted SED is not strongly sensitive to the adopted inner radius, but outer radii much larger than the adopted value result in dust too cool to match the observed SED.

We first computed a set of models for the $+230$\,d epoch (shown in Fig.\,\ref{fig_sndustSEDevolution}), spanning both silicate and carbon dust compositions with grain sizes ranging from 0.1 to 5.0\,$\mu$m. At this epoch, we find that silicate dust is heated to above its assumed sublimation temperature of 1400\,K and thus destroyed unless the grain radius exceeds 1\,$\mu$m. Much smaller grains of amorphous carbon dust can survive due to their higher sublimation temperature of 2200\,K, and their formation at this early epoch is more plausible.
We therefore assume that the dust is composed only of amorphous carbon.

While Rayleigh-Taylor instabilities are expected to form during SN explosions and give rise to clumpy ejecta distributions \citep[e.g.][]{1989ApJ...341L..63A}, the optical non-detections are difficult to reconcile with a clumpy dust distribution. In the case of SN~1987A, clumpy dust models have found that a clump volume filling factor of around 0.1 is required (\citealt{2007MNRAS.375..753E, 2015MNRAS.446.2089W, 2016MNRAS.456.1269B}). If this is adopted for SN~2018bsz, optical emission exceeding the observed non-detections is predicted due to the high probability of photons escaping the ejecta without encountering a dense dust clump. We therefore investigated clumpy dust distributions with much higher volume filling factors, and find that a filling factor greater than about 0.7 is required to match the optical non-detections. For a given total mass, the predicted SED for smooth dust distributions are not significantly different to the predictions for clumpy distributions filling so much of the volume, and so for simplicity, we adopt a smooth dust distribution. With this assumption, we find that the $+230$\,d SED can be well fitted by $5\times 10^{-4}\,M_\odot$ of amorphous carbon dust with a grain size of 0.1\,$\mu$m.

At days $+230$\,d and $+265$\,d, we use the luminosity from the bolometric light curve (Fig.\,~\ref{fig:bol_fra}) in our models. To fit subsequent epochs, we investigated different rates of decline of the luminosity. From its peak until $+265$\,d, the bolometric light curve declines with a half-life of approximately 32 days, much faster than the half-life of $^{56}$Co. However, extrapolating this to $+535$\,d, we find that the luminosity would then be much too low to reproduce the observed IR fluxes. To constrain the luminosity at this epoch, we consider two limiting cases. We first estimate an upper limit to the luminosity of $1.7\times 10^{41}\,\mathrm{erg\,s^{-1}}$; above this, the emission from a 5\,000\,K blackbody would exceed the observed infrared fluxes. We then estimate a lower limit by considering models containing $10^{-3}\,M_\odot$ of dust, which maximises the emission at 3.6 and 4.5\,$\mu$m (higher/lower dust masses give predicted SED which peak at longer/shorter wavelengths). To match the observed fluxes at these wavelengths with this dust mass requires a luminosity of about $1.7\times 10^{40}\,\mathrm{erg\,s^{-1}}$.

We therefore fitted subsequent epochs by expanding the modelled dusty shell, reducing the luminosity according to $^{56}$Co radioactive decay, and then varying the dust mass uniformly throughout the shell if necessary. We note that while the late time light curve of SN~2018bsz is probably not powered by Ni \citep[e.g.][]{2013ApJ...770..128I}, the decline rate of SLSNe is quite similar to that of $^{56}$Co until 600 days \citep{2017ApJ...835..177M}, making this a reasonable choice of function for the input luminosity.

\begin{figure}
\includegraphics[width=0.5\textwidth]{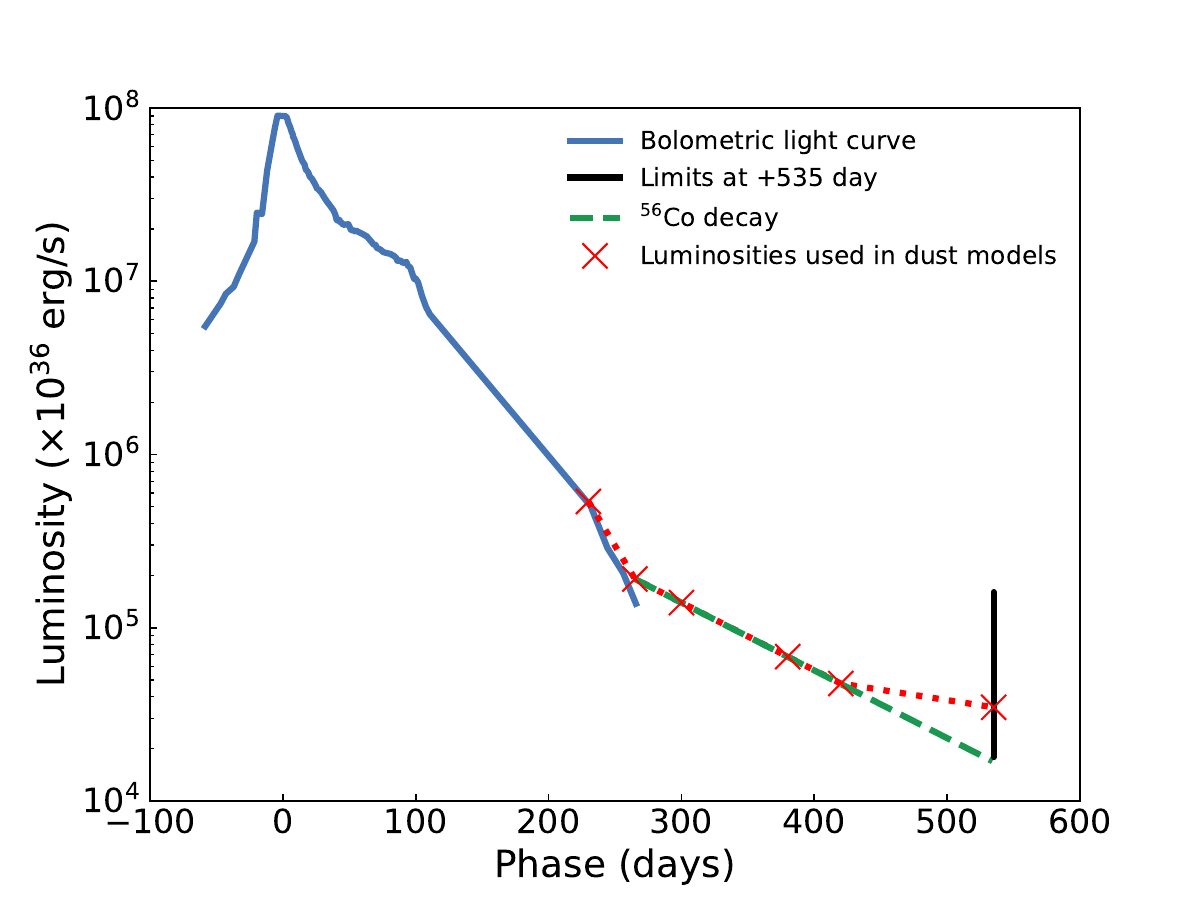}
\caption{The observed bolometric light curve of SN~2018bsz and late time extrapolations. Since we only detect SN~2018bsz in the MIR at +535\,d, we must estimate the true bolometric luminosity that serves as input to our dust models. The lower limit on the late time luminosity (lower edge of the black vertical line) comes from the requirement that the modeled dust mass does not decrease from previous epochs. The upper limit (upper edge of the black vertical line) comes from the most luminous 5,000\,K blackbody fit which matches the MIR points. The green dashed line shows the $^{56}$Co decay, while the orange points show the final luminosities that we use in our dust modeling.}
\label{fig_LSEDevolution}
\end{figure}

\begin{figure*}
\includegraphics[width=\textwidth,trim={0.5cm 0cm 0.5cm 0.5cm}]{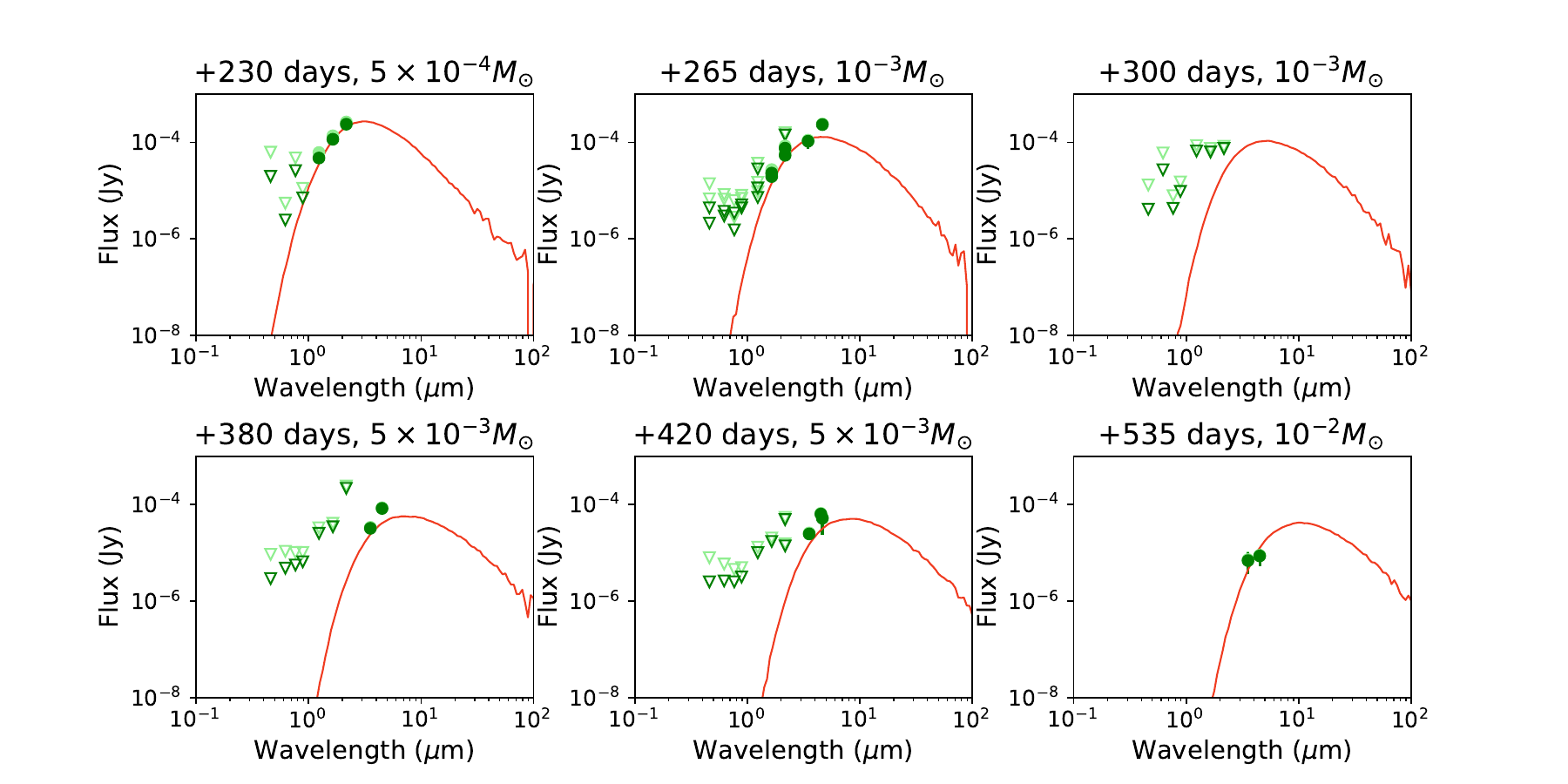}
\caption{Predicted emission from new dust in the SN ejecta at epochs from 
$+230$\,d to $+535$\,d after the peak. The symbols are the same as those used in Fig.\,3. Here, we also include the SN magnitudes assuming a high host‐galaxy extinction of$A_{V}=1$\,mag, shown in light green.}
The assumed dust model is described in Sec.\,\ref{sec_new_dust_ejecta}.
\label{fig_sndustSEDevolution}
\end{figure*}

We find that until $+535$\,d, models with a steadily increasing dust mass can fit the observations. At $+535$\,d, calculating the luminosity by extrapolating the $^{56}$Co decay, we are able to fit the observed IR fluxes but with a lower dust mass than at $+420$\,d, which is unphysical. 
Enforcing a non‐decreasing dust mass therefore requires a luminosity exceeding the $^{56}$Co prediction to fit the MIR emission. 
Figure\,\ref{fig_LSEDevolution} shows the bolometric light curve, the extrapolated $^{56}$Co decline, and the actual luminosities used in our models.

\begin{figure*}
\includegraphics[width=\textwidth,trim={0cm 2.9cm 0cm 0cm}]{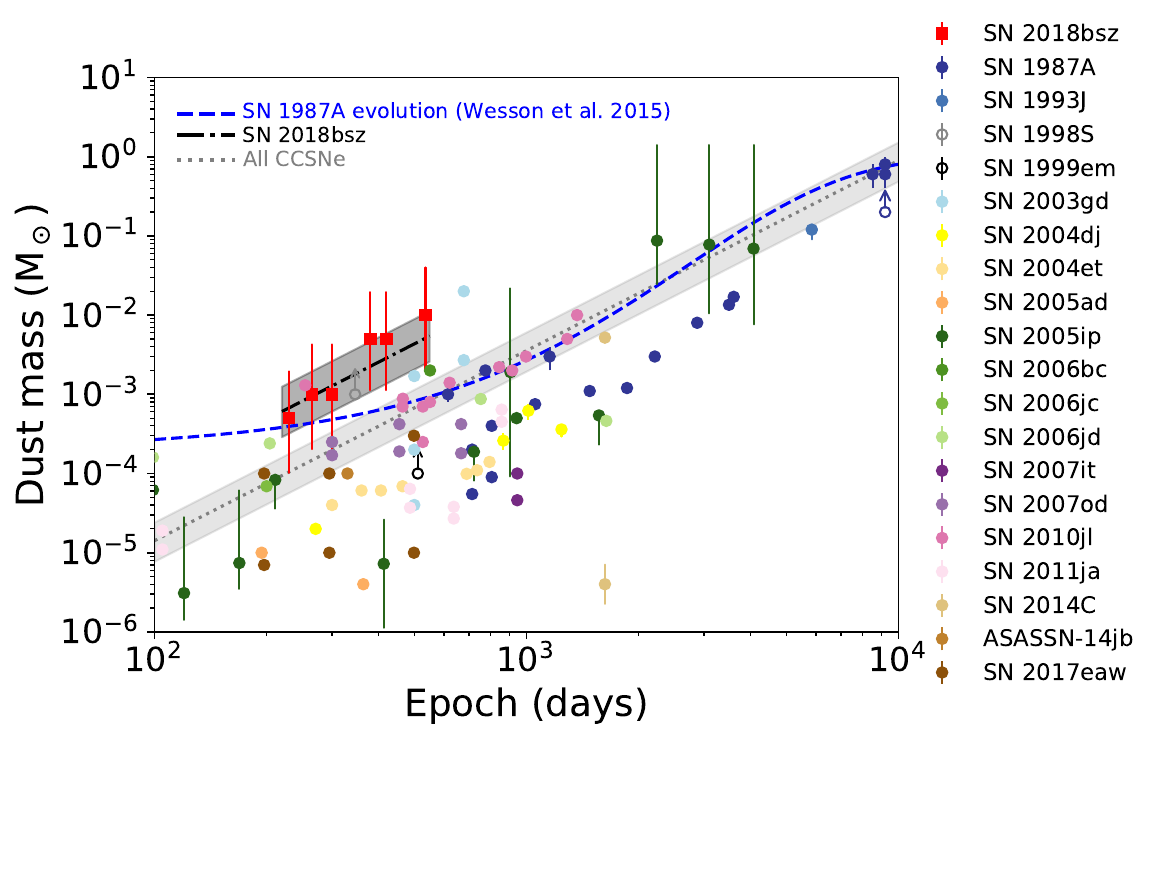}
\caption{The dust masses inferred for SN~2018bsz, compared with literature values for other CCSNe.
We find that SN~2018bsz formed ten times more dust than normal core-collapse SNe at similar epochs.
We fit the evolution of SN~2018bsz and all other CCSNe separately, using M$_{\rm{dust}} \propto t^{2.4}$ (adopting the SN~2010jl ejecta dust evolution from \citealt{2014Natur.511..326G}), and show the evolution in black dash-dotted and grey dotted lines, respectively. The grey areas indicate the 95\% confidence level, incorporating dust mass uncertainties to fit the points for SN~2018bsz and other CCSNe.
The blue dashed line shows the evolution of the dust mass in SN~1987A inferred by \citet{2015MNRAS.446.2089W}. See the webpage\textsuperscript{a} for the full collection; the SNe and references we used in this figure are SN~1987A\textsuperscript{b} \citep{2011Sci...333.1258M, 2014ApJ...782L...2I, 2015ApJ...800...50M, 2015MNRAS.446.2089W, 2016MNRAS.456.1269B}, SN~1993J \citep{2017MNRAS.465.4044B}, SN~1998S \citep{2004MNRAS.352..457P}, SN~1999em \citep{2003MNRAS.338..939E}, SN~2003gd \citep{2006Sci...313..196S, 2007ApJ...665..608M}, SN~2004dj \citep{2011A&A...527A..61S}, SN~2004et \citep{2009ApJ...704..306K}, SN~2005ad and SN~2005af \citep{2013A&A...549A..79S}, SN~2005ip \citep{2010ApJ...725.1768F, 2012ApJ...756..173S, 2019MNRAS.485.5192B}, SN~2006bc \citep{2012ApJ...753..109G}, SN~2006jc \citep{2009ApJ...692..546S}, SN~2006jd \citep{2012ApJ...756..173S}, SN~2007it \citep{2011ApJ...731...47A}, SN~2007od \citep{2010ApJ...715..541A}, SN~2010jl\textsuperscript{c} \citep{2013ApJ...776....5M, 2018ApJ...859...66S, 2020ApJ...894..111B}, SN~2011ja \citep{2016MNRAS.457.3241A}, SN~2014C \citep{2019ApJ...887...75T}, ASASSN-14jb \citep{2019A&A...629A..57M}, and SN~2017eaw \citep{2019ApJ...873..127T}.}

\textsuperscript{a:} {\scriptsize https://www.nebulousresearch.org/dustmasses/}

\textsuperscript{b:} {\scriptsize
We note that \citet{2015ApJ...810...75D} report high dust masses at early epochs for SN~1987A. However, we do not plot these here as they are incompatible with the observed gradual evolution of the emission-line profiles \citep{2021ApJ...923..148W}.}

\textsuperscript{c:} {\scriptsize
For SN~2010jl we show the dust masses assuming carbon dust in \citet{2018ApJ...859...66S}.}
\label{fig_dustcomparison}
\end{figure*}

Our predicted SED at each of the six epochs considered is shown in Fig.\,\ref{fig_sndustSEDevolution}. 
We emphasise that the model fits are constrained not only by the data at each epoch, but also the requirement that the model must evolve self-consistently between epochs. In our best-fitting set of models, the dust mass increases from about $5\times 10^{-4}\,M_\odot$ at $+230$\,d to $10^{-2}\,M_\odot$ at $+535$\,d. This pattern of the dust mass increasing steadily over a few hundred days is consistent with the pattern of dust formation observationally determined in other CCSNe \citep[e.g.][]{2015MNRAS.446.2089W, 2016MNRAS.456.1269B, 2020ApJ...894..111B}. In Fig.\,\ref{fig_dustcomparison} we plot literature values of estimated dust masses at different epochs of other CCSNe (between $10^{2}$ and $10^{4}$ days) together with our estimated masses for the dust in SN~2018bsz. 
At all epochs, the largest source of uncertainty in the estimate of the dust mass is the adopted luminosity. At $+230$ and $+265$\,d, the bolometric luminosity is well-constrained, and the resulting uncertainty in the dust mass is about a factor of 2. At $+300$\,d where only upper limits to the fluxes are available, the dust mass is unconstrained. At later epochs when the luminosity is extrapolated, the uncertainties are larger. At $+535$\,d, as described above, a dust mass of zero is nominally compatible with the observations but would require a much higher luminosity than at earlier epochs. To estimate the uncertainty on the dust mass at this epoch, we require that the luminosity is not higher than at previous epochs, and then vary both the dust mass and luminosity to obtain fits to the SED. In this way we estimate that the dust mass uncertainty is about a factor of five. The possible contribution of a CSM echo at this epoch may introduce an additional uncertainty, although if we assume that the mass of newly-formed ejecta dust does not decrease at any point, the overestimate in the dust mass that would be caused by attributing a CSM echo to newly-formed dust emission must be small.

These uncertainties notwithstanding, SN~2018bsz appears to have formed more dust at these epochs than most other CCSNe. 
The shape of the dust growth curve of SN~2018bsz is very similar to the general trend seen in Fig.\,\ref{fig_dustcomparison}, and SN~2018bsz clearly lies at the upper edge of the spread seen in normal CCSNe. We fit the evolution of SN~2018bsz and all other CCSNe separately using M$_{\rm{dust}} \propto t^{2.4}$ (adopting from the SN~2010jl ejecta dust evolution from \citealt{2014Natur.511..326G}), and find SN~2018bsz to have formed ten times more dust than normal CCSNe at similar epochs. 
In the case of SN~1987A, \citet{2015MNRAS.446.2089W} found that the majority of the dust present 23 years after the explosion must have formed at epochs later than 1000 days. If a similar formation process is occurring in SN~2018bsz, the $10^{-2}\,M_\odot$ of dust present at $+535$\,d will be a small fraction of the final dust mass.

Finally, even for the largest plausible value of host galaxy extinction $A_V \sim 1$\,mag, the corresponding MIR extinctions in the Spitzer/IRAC bands at 3.6\,$\mu$m (channel 1) and 4.5\,$\mu$m (channel 2), denoted as $A_{ch1}$ and $A_{ch2}$, are only 0.07\,mag and 0.05\,mag, respectively, and can therefore be neglected. The foreground dust screen and light echo models are still incompatible with our observations. The model for newly formed dust in the ejecta is, however, still consistent with the data (see Fig.\,\ref{fig_sndustSEDevolution}).

\section{Discussions and conclusion}
\label{sec:discus}

Several lines of evidence suggest that new dust forming in the ejecta of SN~2018bsz is responsible for the observed IR excess after $\sim200$ days.
CSM echoes cannot reproduce the temporal evolution of SN~2018bsz; while the CSM geometries that could explain the IR emission at much later times ($>230$ days) would also predict detectable optical emission from scattered light, which we do not observe. ISM echoes, on the other hand, would have to be from extremely dense dust clouds close to the SN to give rise to detectable IR emission. We find that a sheet of dust 1\,pc in front of the SN and 0.1\,pc deep, containing $1\,M_\odot$ of dust, would result in an echo at $535$\,d with fluxes about two orders of magnitude below those observed. 
Therefore, our modelling suggests newly-formed dust in the SN ejecta at epochs $>200$\,days, with a dust composition of pure amorphous carbon. 
We estimate that the dust mass formed in SN~2018bsz is about ten times larger than what is found for other SNe at similar phases, with the dust mass increasing from about $5\times 10^{-4}\,M_\odot$ at $+230$\,d to $10^{-2}\,M_\odot$ at $+535$\,d. 

While our models reproduce the optical non-detections, infrared excess and overall evolution of the SED, they do not perfectly match the data. Most obviously, fluxes at 4.5\,$\mu$m are underpredicted by the models at +265, +380 and +420~days. This may be due to emission from the CO fundamental band at 4.6\,$\mu$m, which is not included in our models; an excess at this wavelength has been observed at early times in other supernovae and attributed to CO (e.g., in SN~2011dh; \citealt{2014A&A...562A..17E}). In the absence of spectroscopic data at these epochs, we cannot confirm the origin of the 4.5\,$\mu$m excess.

Better fits to the SED at all epochs might also be possible with a more sophisticated model invoking a distribution of grain radii instead of a single grain size. However, this would add several more free parameters to the model which the available data could not uniquely constrain. Similarly, given that silicate dust could not be present at the earlier epochs unless the grains were already extremely large, we adopt pure carbon at all epochs. Silicate dust formation at later epochs when the ejectas has cooled sufficiently is not ruled out, but is not included in our models.

However, we stress that scenarios invoking thermal echoes from ISM or CSM dust cannot be reconciled with the data, being ruled out both by their predicted temporal evolution, and by optical non-detections to several orders of magnitude fainter than the fluxes predicted by these models. We hence suggest that newly formed dust is the only credible explanation for the late time IR emission that is seen in SN~2018bsz.

The fact that SN~2018bsz was hydrogen poor at early times \citep{2018A&A...620A..67A}, and that the Balmer lines appear after 77 days \citep{2022A&A...666A..30P}, before dust formation (after 200 days), suggests that the dust formation may be enhanced by the interaction between the SN ejecta and a hydrogen-rich CSM.
Mixing of metal-rich material in the ejecta with a hydrogen-rich CSM can lead to the formation of large amounts of dust, as the necessary molecular processes are thought to be more efficient in a hydrogen-rich environment \citep{1990MNRAS.246..208R}. 

Late-time CSM interaction, together with a strongly asymmetric ejecta (as seen, for example in Luminous Blue Variables like Eta Carinae) may also account for the peculiar Balmer line profiles. In this scenario, around 100 days after peak brightness we observe the Balmer lines first forming, and later when the SN ejecta sweeps up the CSM it causes the observed changes in their profiles. The blue side of the line profiles is absorbed by swept-up CSM, while thanks to the asymmetric geometry, we can still see the red side of the lines. We do not see such profiles in the metal lines, likely because they come from deeper layers in the ejecta. Dust will form as the shocked, swept-up CSM cools, as is observed at late times ($>200$ days).

Although SN~2018bsz is the first SLSN without hydrogen lines at maximum light for which dust has been detected, a late-time NIR excess was observed in one Type II SLSN, SN~2008es ($M_V\sim-22$\,mag)
\citep{2009ApJ...690.1313G, 2009ApJ...690.1303M, 2018MNRAS.475.1046I}. 
It has a broad H$\alpha$ feature during the photospheric phase and does not show any sign of strong interaction between SN ejecta and circumstellar shells in the early spectra. 
\citet{2019MNRAS.488.3783B} obtained late-time photometry of SN~2008es, which showed $r-K'>$1.5\,mag around 250--300 days after the explosion. The authors suggested that this NIR excess indicates dust condensation in a cool dense shell that formed through circumstellar interaction \citep[e.g.][]{2008MNRAS.389..141M}.
We note however that SN~2018bsz has a much redder colour of $r-K_{\mathrm{s}}>4.9$\,mag at a similar epoch to SN~2008es. 

\citet{2019MNRAS.484.5468O} investigated the effect of a pulsar wind
nebula on dust formation and found that the pulsar wind can accelerate or delay dust formation. The formation timescale of carbon dust can vary from $\sim180$ to $\sim590$ days for Type I SLSN assuming a magnetar power source. 
In our case for SN~2018bsz we detect carbon dust formation within one year after explosion, which is consistent with their predicted timescale.

Type I SLSNe are rare events, with about one SLSN for every 3500$^{+2800}_{-720}$ CCSNe estimated from the Palomar Transient Factory sample by \citet{2021MNRAS.500.5142F}. 
These authors found that the relative rate was higher in low-mass ($\log(M_\star/M_\odot)<9.5$) galaxies, with one SLSN for every $1700^{+1800}_{-720}$ CCSNe.
Furthermore, sub-solar metallicity appears to be required for massive stars to explode as Type I SLSNe \citep{2016ApJ...830...13P, 2017A&A...602A...9C, 2018MNRAS.473.1258S}. Such low-mass, low-metallicity galaxies are common in the early Universe  (\eg{}\citealt{Heckman+05,Cardamone+09,Bian+16}). 
The dust mass deduced for SN~2018bsz is $\sim10$ times larger than what is seen in most CCSNe. 
If the rate of Type I SLSNe were $10^{-2}$ of the CCSN rate in the high-$z$ Universe, they could thus produce up to 10\% of the dust in the early Universe if this dust is due to CCSNe only. In fact, \citet{2019ApJS..241...16M} observed SLSNe at high redshifts and determined that their volumetric rate is 5 ($3.5 \lesssim z \lesssim 4.5$) to 10 ($1.5 \lesssim z \lesssim 2.5$) times higher than the local SLSN rate. One could even speculate that if SLSNe are produced by more massive stars than CCSNe, and if the initial mass function at high redshift is top heavy, then SLSNe could be even more common.

A more self-consistent (although still approximate) estimate of the dust mass produced by Type I SLSNe at high redshift can be obtained from cosmological galaxy evolution simulations. For example, the cosmic dust production rate density (DD) from SLSNe can be calculated from the latest version of the \textsc{L-Galaxies} semi-analytic simulation \citep{Yates+24}, which contains self-consistent models for both metal production by stars and SNe and dust production and destruction. In \textsc{L-Galaxies}, the net dust mass ejected by CCSNe is assumed to be proportional to the ejected mass of the key chemical element from which it forms (\ie{}C, Mg, Si, or Fe, see \citealt{2019MNRAS.489.4072V} section 3.1). The early destruction of newly-condensed dust by the reverse SN shock is accounted for in the dust model via a net `condensation efficiency parameter', which is evaluated from dust destruction models (see \citealt{2008A&A...479..453Z}). The later destruction of dust in the ISM via SN shock wave propagation is also considered. The element mass ejection rate is calculated using a galactic chemical enrichment (GCE) scheme, which tracks the birth of stellar populations and the subsequent ejection of heavy elements by a range of SN types and stellar wind mechanisms.

For the analysis here, we assume that (a)  Type I SLSNe only form in galaxies with gas-phase metallicity $\leq 0.5\, Z_\odot$, (b) SLSNe produce 10 times as much dust per event as CCSNe, (c) the CCSNe/SLSNe rate ratio is 3500 at $z=0$, and (d) the CCSN rate is proportional to the cosmic SFR density (which matches that observed by \citealt{2014ARA&A..52..415M} in \textsc{L-Galaxies} at high redshift, see \citealt{Yates+21b} section 4.2). Under these assumptions, \textsc{L-Galaxies} predicts a SLSN DD of log$_{10}(M_{\rm dust,SLSN}$ / M$_\odot$\,yr$^{-1}$\, Mpc$^{-3}) \sim{} -7.14$ at $z=7$. This is around two orders of magnitude lower than the DD expected from CCSNe at $z=7$ (see \citealt{Yates+24}, fig. 8). However, it is higher than the contribution expected from SNe-Ia and only an order of magnitude below that expected from AGB stars, which themselves could play an important role (see \citealt{2009MNRAS.397.1661V,Mancini+15}). Therefore, SLSNe could provide a non-negligible contribution to dust production in the high-redshift Universe.

We caution that these estimates assume that all SLSNe produce as much dust as SN~2018bsz. However, SN~2018bsz is a somewhat unusual SLSN and appears quite different to the bulk of the Type I SLSN population (according to its unique light curve-shape and spectroscopic evolution).
It is hence quite possible that only a few percent of the SLSN population produce dust.

It should nevertheless be pointed out that observational data for Type I SLSNe remains the limiting factor in knowing their dust masses. We may well be systematically missing the window to see evidence for dust-formation in SLSNe, as the majority of them are found at $z>0.1$ and do not have rest-frame $K$-band observations, especially at late times ($>200$ days). In fact, only two other Type I SLSNe (among $\sim$200 events) have late-time rest-frame K-band data, namely PTF12dam \citep{2013Natur.502..346N,2015MNRAS.452.1567C,2017ApJ...835...58V}, and SN~2015bn \citep{2016ApJ...826...39N},
although some SLSNe have displayed a rebrightening at mid-infrared wavelengths after 200 days \citep{2022MNRAS.513.4057S}. 
A concerted effort to acquire more late-time IR data is therefore needed in order to gain a greater understanding of the contribution from Type I SLSNe in the early Universe.

Finally, a high luminosity combined with their preference for dwarf hosts, makes Type I SLSNe appealing targets for high-redshift (young Universe) searches, e.g. with \textit{Euclid} \citep{2018A&A...609A..83I}, the \textit{James Webb Space Telescope} (\textit{JWST}) or the Rubin Observatory Legacy Survey of Space and Time (LSST) \citep{2021MNRAS.504.2535I}. 
\textit {JWST} in particular can reach down to 27th magnitude in the infrared (IR), potentially allowing for observations of SLSN out to $z=20$ \citep{2019BAAS...51c.399W}, and spectroscopic classification of some of these up to $z\sim5$ \citep{2020NatAs...4..893N}.
Further systematic observations by \textit{JWST} will help to determine how much dust is formed in typical Type I SLSNe.

\acknowledgments
We gratefully acknowledge the late Patricia~Schady for her important contributions to this work, and we deeply appreciate her insight, support, and collaboration.
We thank Prof. Michael~J.~Barlow for his valuable discussions and inputs about ejecta dust and circumstellar dust. We would also like to thank Margherita Molaro for valuable discussions on the analytic estimation of dust masses at high redshift.
We also thank the anonymous referee for their careful reading of the paper and comments.
T.-W.~C. thanks Sijie~Chen, Dan~Perley, Jinyi~Shangguan, Lin~Yan, Jason~Spyromilio, Stefan~Taubenberger, Seppo~Mattila, and John~Danziger for useful discussions of this object.

T.-W.~C. acknowledges funding from the Alexander von Humboldt Foundation, from the EU under the Marie Sk\l{}odowska-Curie Individual Fellowship (H2020-MSCA-IF-2018-842471), from the Yushan Fellow Program of the Ministry of Education (MOE-111-YSFMS-0008-001-P1), and from the National Science and Technology Council, Taiwan (NSTC 114-2112-M-008-021-MY3).
S.~J.~B. acknowledges support from Science Foundation Ireland and the Royal Society (RS-EA/3471).
R.~W. acknowledges support from European Research Council (ERC) Advanced Grant 694520 SNDUST (PI: Barlow). 
A.~J.~R. is thankful for support from the Australian Research Council under award number FT170100243.
M.~F. is supported by a Royal Society - Science Foundation Ireland University Research Fellowship.
E.~Y.~H. acknowledges the support provided by the National Science Foundation under Grant No. AST-1613472 and by the Florida Space Grant Consortium.
C.-J.~L. is supported by the NSTC grant 114-2112-M-004-001-MY2 from the National Science and Technology Council of Taiwan, and the MOST grant 109-2112-M-001-040 and 109-2811-M-001-545 from the Taiwanese Ministry of Science and Technology.
S.~S. is partially supported by LBNL Subcontract 7707915.
M.~N. is supported by the European Research Council (ERC) under the European Union's Horizon 2020 research and innovation programme (grant agreement No.~948381) and by UK Space Agency Grant No.~ST/Y000692/1.
G.~L. was supported by a research grants (19054 and 60862) from VILLUM FONDEN.
A.~G.-Y.'s research is supported by the EU via ERC grant No. 725161, the ISF GW excellence center, an IMOS space infrastructure grant and BSF/Transformative and GIF grants, as well as The Benoziyo Endowment Fund for the Advancement of Science, the Deloro Institute for Advanced Research in Space and Optics, The Veronika A. Rabl Physics Discretionary Fund, Minerva, Yeda-Sela and the Schwartz/Reisman Collaborative Science Program; A.~G.-Y. is the recipient of the Helen and Martin Kimmel Award for Innovative Investigation.
L.~G. acknowledges financial support from AGAUR, CSIC, MCIN and AEI 10.13039/501100011033 under projects PID2023-151307NB-I00, PIE 20215AT016, CEX2020-001058-M, ILINK23001, COOPB2304, and 2021-SGR-01270.
C.~P.~G. acknowledge financial support from AGAUR, CSIC, MCIN and AEI 10.13039/501100011033 under projects PID2023-151307NB-I00, PIE 20215AT016, CEX2020-001058-M, ILINK23001, COOPB2304, and 2021-SGR-01270; from the Secretary of Universities and Research (Government of Catalonia) and by the Horizon 2020 Research and Innovation Programme of the European Union under the Marie Sk\l{}odowska-Curie and the Beatriu de Pin\'os 2021 BP 00168 programme.
M.~G. is supported by the EU Horizon 2020 research and innovation programme under grant agreement No 101004719.
S.~G.~G. acknowledges support by FCT under Project CRISP PTDC/FIS-AST-31546/2017 and UIDB/00099/2020.
K.~M. acknowledges funding from the European Union (ERC, CosmicLeap, 101125877).
G.~P. is supported  by ANID – Millennium Science Initiative - ICN12\_009.
M.~P. acknowledges support from a UK Research and Innovation Fellowship(MR/T020784/1)
H.~K. was funded by the Research Council of Finland projects 324504, 328898, and 353019.

GROND observations at La Silla were performed as part of programs 102.A-9099 and 103.A-9099. Part of the funding for GROND (both hardware as well as personnel) was generously granted from the Leibniz-Prize to Prof. G. Hasinger (DFG grant HA 1850/28-1). 
This work is based (in part) on observations collected at the European Organisation for Astronomical Research in the Southern Hemisphere, Chile as part of ePESSTO and ePESSTO+, (the Public ESO Spectroscopic Survey for Transient Objects Survey), ESO programs 199.D-0143 and 1103.D-0328.
We also collect data from the ESO Science Archive Facility using the SINFONI instrument under the ESO program 2101.D-5026, the X-Shooter instrument under ESO program 0103.D-0697 and the MUSE instrument under ESO program 0103.D-0523.
This work is based [in part] on observations made with the \textit{Spitzer Space Telescope}, which was operated by the Jet Propulsion Laboratory, California Institute of Technology under contract with NASA, \textit{Spitzer} Observing program ID: 14273. 
This publication makes use of data products from the \textit{Wide-field Infrared Survey Explorer}, which is a joint project of the University of California, Los Angeles, and the Jet Propulsion Laboratory/California Institute of Technology, funded by the National Aeronautics and Space Administration.
We acknowledge the use of public data from the Neil Gehrels \emph{Swift} Observatory data archive, under target-of-opportunity programs . 

The {\sc STARLIGHT} project is supported by the Brazilian agencies CNPq, CAPES and FAPESP and by the France-Brazil CAPES/Cofecub program.
{\sc IRAF} is distributed by the National Optical Astronomy Observatory, which is operated by the Association of Universities for Research in Astronomy (AURA) under a cooperative agreement with the National Science Foundation.
This research made use of {\sc PypeIt}, a Python package for semi-automated reduction of astronomical slit-based spectroscopy.

\clearpage

\appendix

\renewcommand\thefigure{A\arabic{figure}}    
\setcounter{figure}{0} 

\section{Imaging observations of SN~2018bsz}
\label{sec:app_images}

\subsection{UVOT observations with Swift}
\label{sec:app_imaging_swift}

We obtained 28 epochs of photometry for SN~2018bsz with UVOT on board the Neil Gehrels \emph{Swift} Observatory. 
The UVOT photometry was performed using the task \texttt{uvotsource} in HEASoft\footnote{\href{https://heasarc.gsfc.nasa.gov/docs/software/heasoft/}{https://heasarc.gsfc.nasa.gov/docs/software/heasoft/}} version 6.25, with a 3\arcsec-radius aperture. For host subtraction, a template was constructed from images obtained between MJD = 58562 and 59068. We measured the host contribution using the same source and background apertures and numerically subtracted the host contribution from the SN light curve.
While this manuscript was in preparation, a new version of the Swift CALDB was released which affects the zeropoint for our UVOT data. However, as the change in photometry is less than 0.1 mag and within the photometric uncertainties in all epochs, we have not implemented this change here, and it will not affect our results.
Table\,\ref{tab:phot_swift} provides the UVOT host-subtracted light curve.

\begin{figure}
    \centering
    \includegraphics[width=0.8\textwidth]{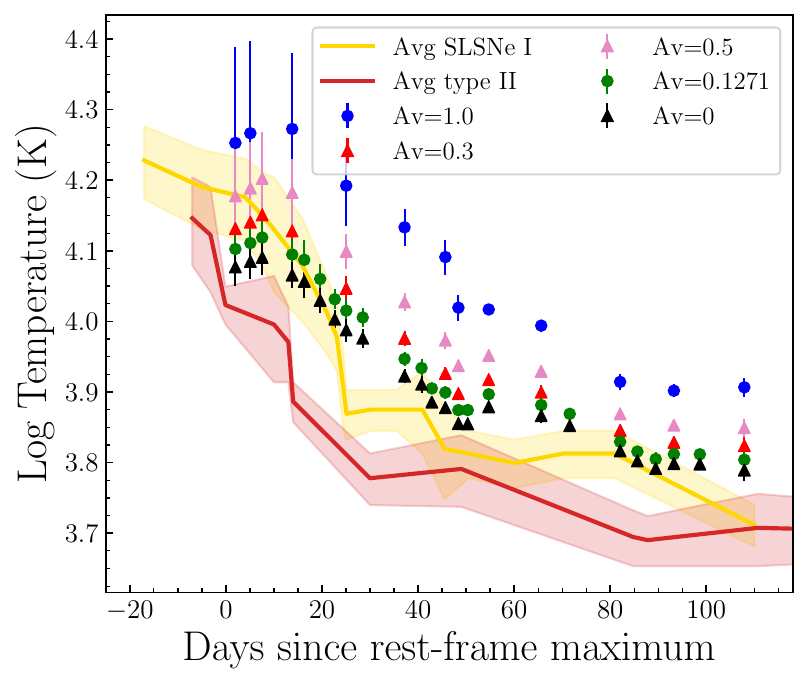}
    \caption{Temperature evolution of SN~2018bsz in comparison to the average evolution of Type I SLSNe and Type II SNe \citep{2018MNRAS.475.1046I}. The comparison samples comprise 14 Type I SLSNe \citep{2013ApJ...770..128I, 2015MNRAS.452.3869N} and $\sim30$ Type II SNe \citep{2003MNRAS.338..939E, 2010MNRAS.404..981M, 2011MNRAS.417..261I, 2012MNRAS.422.1178I, 2010ApJ...717L..52B, 2013MNRAS.434.1636T, 2017ApJ...850...90G}. The shaded regions represent the 99.73 percent confidence ranges for temperature of each class of SNe. The temperature of SN~2018bsz is measured from SED fits assuming a blackbody at selected epochs. We find that if the host extinction is between $A_{V}=0.1$ and 0.3 mag we obtain reasonable agreement with the SLSNe at 3--20 days after the peak.}
    \label{fig:BB_temp}
\end{figure}

\subsection{Optical and near-infrared imaging from the ground}
\label{sec:app_imaging_ground}

We used GROND to monitor the light-curve evolution of SN~2018bsz from $+3$\,d to $+421$\,d. This 7-channel imager collects multi-band photometry simultaneously in the \textit{g', r', i', z', J, H} and $K_{\mathrm{s}}$ bands.
The images were reduced using the GROND pipeline \citep{2008ApJ...685..376K}, which applies de-bias and flat-field corrections, stacks images and provides astrometric calibration. For GROND NIR images, due to the bright host galaxy we disabled line by line fitting of the sky subtraction routine since this caused over-subtraction artifacts. 

All photometric measurements of the imaging data were carried out using the custom-built photometry pipeline {\sc AutoPhOT}\footnote{https://github.com/Astro-Sean/autophot} \citep{2022A&A...667A..62B}; the version of the software used in this work is archived on Zenodo \citep{brennan_2026_21534239}. Point spread function (PSF) photometry was performed using a PSF model built from bright, isolated sources in the image. We adopted the $+610$\,d epoch as the host galaxy template and subtracted that before measuring the SN brightness using {\sc AutoPhOT}. The optical magnitudes are calibrated against PanSTARRS field stars while NIR images are calibrated to 2MASS and converted to the AB system. The PSF model for each image is used to determine limiting magnitudes. We collect a random sample of pixels within a cutout around SN~2018bsz and build a pseudo-PSF model. We then assert a confidence that pixels at the transient are real detection, rather than a correlated noise spike.

\begin{figure*}
    \centering
    \includegraphics[width=0.8\textwidth]{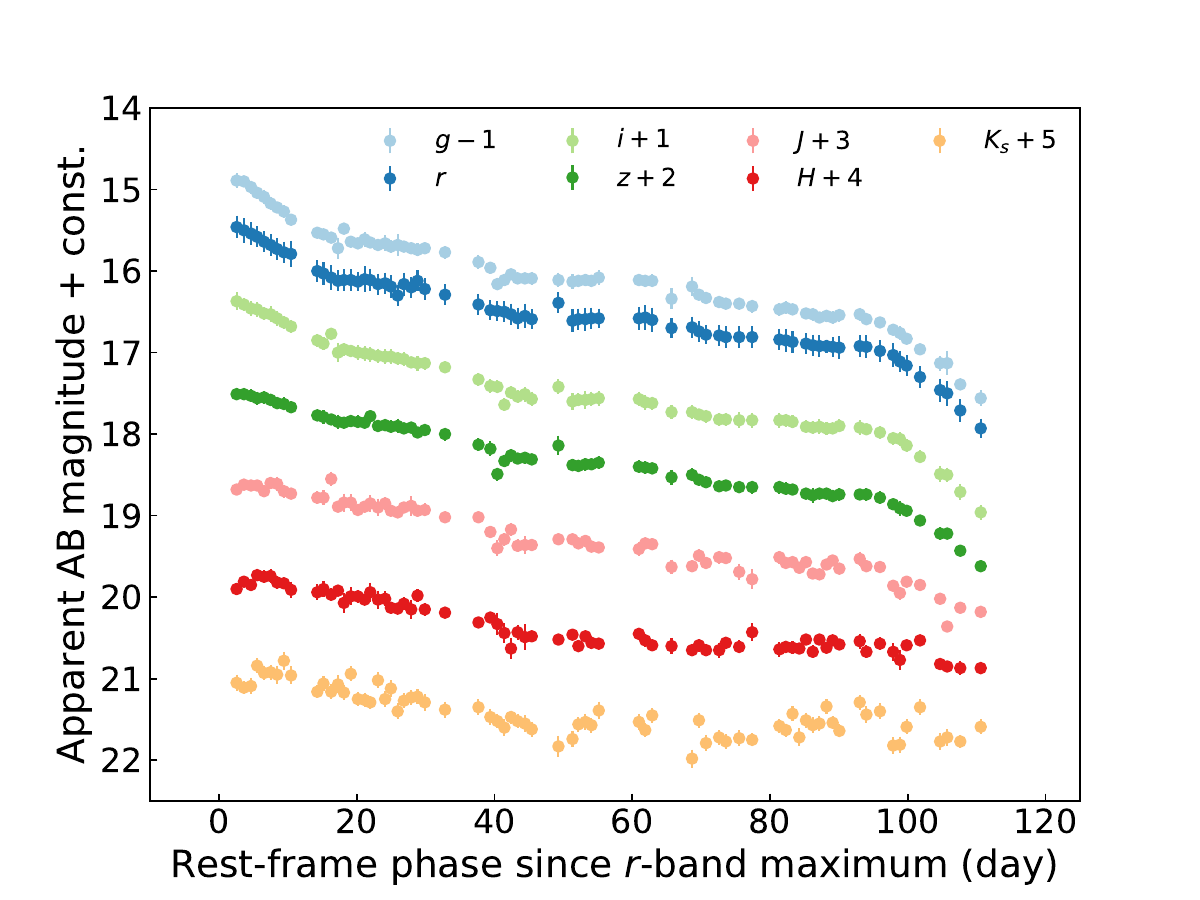}
    \caption{GROND host-subtracted light curves of SN~2018bsz. These data are the first daily-cadence and simultaneous 7-band observations of an SLSN.}
    \label{fig:grond_phot_roomin}
\end{figure*}

For this work, limiting magnitudes are set to $3\sigma$ confidence. For further details, see \citet{2022MNRAS.513.5642B}.
Figure\,\ref{fig:grond_phot_roomin} shows our GROND light curves of SN~2018bsz from $+3$\,d to $ +111$\,d.

In addition, we obtained several \textit{J} and $K_{\mathrm{s}}$-band images using SOFI on the NTT through the ePESSTO program. The images were reduced using the PESSTO pipeline \citep{2015A&A...579A..40S} with flat-fielding and sky background subtraction. The host templates were taken around $+600$\,d and we measured the host-subtracted photometry of those SOFI images using {\sc AutoPhOT}. 

The optical and NIR host-subtracted photometry of SN~2018bsz is reported in Tables \ref{tab:phot_grond_optical}, \ref{tab:phot_grond_nir} and \ref{tab:phot_sofi}.

\subsection{Mid-infrared imaging from space}
\label{sec:app_imaging_space}

We obtained \textit{Spitzer} observations at four different epochs between 2019 June 27 (at epoch $+384$\,d) and 2019 December 30 ($+564$\,d), just before the end of the \textit{Spitzer} mission. 
The images were taken using IRAC in the 3.6 and 4.5\,$\mu$m bands, which we here refer to as the $Ch1$ and $Ch2$ bands.

We aligned the images using the {\sc geomap} and {\sc geotran} tasks within {\sc iraf} and employed the High Order Transform of PSF ANd Template Subtraction (hotpants)\footnote{http://www.astro.washington.edu/users/becker/v2.0/hotpants.html} package to carry out the host template subtraction. We used the last epoch as the host template and subtracted the other frames from it. The image subtraction results are shown in Fig.\,\ref{fig:cutouts}. 
We measured the counts from the SN in those host-subtracted images using the {\sc daophot} task within {\sc iraf}. With an aperture size of 6.1 pixel radius, a conversion factor of $8.461595\times 10^{-6}$ and an aperture correction of 1.112, we calculated the SN fluxes in Janskys and converted them to magnitudes. 
More details on conversion factor and aperture correction can be found from the \textit{Spitzer} Heritage Archive\footnote{https://irsa.ipac.caltech.edu/data/SPITZER/docs/irac/iracinstrumenthandbook/24/}$^{,}$\footnote{https://irsa.ipac.caltech.edu/data/SPITZER/docs/irac/iracinstrumenthandbook/27/\#\_Toc410728317}. The errors were dominated by the standard deviation of the sky background pixels around the SN positions.

The \textit{Spitzer} $m_{Ch1}$ measurement at $+384$\,d is consistent with the evolution of the NEOWISE $m_{W1}$ magnitudes. This validates our choice to use the $+565$\,d epoch as a template, because the NEOWISE magnitudes are obtained by subtracting a template that was obtained before the SN explosion. On the other hand, assuming a linear decline from $+384$\,d to $+403$\,d, the extrapolated magnitudes that we derived at $+535$\,d are $m_{Ch1}\sim22.4$, $m_{Ch2}\sim21.4$\,mag, respectively. Those are consistent within the uncertainties of the measured \textit{Spitzer} magnitudes that result from using the $+565$\,d data as the template ($m_{Ch1}=21.8\pm0.7$, $m_{Ch2}=21.6\pm0.6$\,mag).

The NEOWISE-R data set consists of 12 epochs at $W1$ (3.4~$\mu$m) and $W2$ (4.6~$\mu$m) at a $\sim$6 month cadence, with the final three epochs obtained after the discovery of SN~2018bsz. Each epoch consists of 12-18 individual exposures across $\sim$2 days, with PSF-fitted magnitudes available in the NEOWISE-R catalogue. An unresolved source was reported at the position of the host of SN~2018bsz at all epochs. We calculated the average magnitude at each epoch, where exposures flagged for poor quality were excluded, and the error was taken as the standard error of the mean combined with a flux error term to reflect the uncertainty between epochs \citep[e.g.,][]{2020MNRAS.498.2167K}. The nine NEOWISE-R epochs up until the SN discovery did not show significant evolution within the errors at the $W1$ and $W2$ bands. Based on these nine epochs we determined mean quiescent magnitudes in the $W1$ and $W2$ bands of the host of SN~2018bsz of $13.50\pm0.01$ and $13.32\pm0.02$ mag, respectively. The mean quiescent magnitudes were then subtracted, in flux space, from the three post-SN NEOWISE-R epochs to obtain the contribution of the SN in the $W1$ and $W2$ bands at each epoch. The final SN magnitude errors include the statistical error and the systematic error of the quiescent magnitude. More details on this method are described in \citet{2020MNRAS.498.2167K}.

\section{Optical and NIR spectroscopic observations of SN~2018bsz}
\label{sec:app_spectroscopy}

We obtained a series of spectra of SN~2018bsz using different instruments (a log of the observations is given in Table\,\ref{tab:log_spec}). 

SINFONI is a NIR IFU fed by an adaptive optics (AO) module. We used the SN itself as a natural guide star, in order to perform AO corrections. Unfortunately, as SN~2018bsz was around $\sim16$\,mag at the time these observations were taken it was too faint to obtain a satisfactory AO correction. The observations hence have a seeing of $\sim1\arcsec$, much larger than expected.

WiFeS is an IFU with 25 slitlets that are 1\arcsec\,wide and 38\arcsec\,long. The data were reduced using a custom-built pipeline PyWiFeS \citep{2014Ap&SS.349..617C}. The pre-peak NTT+EFOSC2 spectrum was taken from \citet{2018A&A...620A..67A}; the post-peak spectrum at $+108$\,d was reduced in the standard fashion using the PESSTO pipeline \citep{2015A&A...579A..40S}.

The X-Shooter data were reduced using {\sc PypeIt}\footnote{https://pypeit.readthedocs.io/en/latest/} package \citep{2020JOSS....5.2308P, 2020zndo...3743493P}.
However, no SN trace is detected, and in particular we see no clear broad feature at the red extremity of the spectrum where we expect CO emission ($\sim 23,000$\,\AA).

At $+417$\,d, when SN~2018bsz had disappeared in the optical we used MUSE to observe the host environment. 
The data were reduced and combined into single MUSE data cubes using the standard ESO pipeline software version 1.2.11 \citep{2014ASPC..485..451W}. We also subtracted the sky lines from our data cubes using the Zurich Atmospheric Purge \citep[ZAP; v2.0 ][]{2016MNRAS.458.3210S} software and corrected for telluric absorptions using the molecfit software package \citep{2015A&A...576A..77S}. Detailed data reduction procedures are described in \citet{2019MNRAS.490.4515S}. 

We measured the emission line fluxes from the extracted MUSE spectrum at the SN position. The spectrum has been corrected for the Milky Way dust extinction and host dust extinction based on the Balmer decrement, and the stellar continuum has been subtracted. The flux and uncertainty measurement follow the method in \citet{2017A&A...602A...9C}.  
We adopted the \citet{2004MNRAS.348L..59P} calibration of the N2 and O3N2 methods, which use the log([\ion{N}{2}]$\lambda$6583/H$\alpha$) ratio, giving $12+{\rm log(O/H)}=8.27\pm0.06$, and the log([\ion{O}{3}]$\lambda$5007/H$\beta$)/([\ion{N}{2}]$\lambda$6583/H$\alpha$) ratio, giving $12+{\rm log(O/H)}=8.27\pm0.05$, respectively. Moreover, we also adopted a metallicity diagnostic of \cite{2016Ap&SS.361...61D}, which uses [\ion{N}{2}]$\lambda$6583, H$\alpha$, [\ion{S}{2}]$\lambda\lambda$6717,6731 lines, giving $12+{\rm log(O/H)}=8.14\pm0.10$. Assuming a solar oxygen abundance of $12+{\rm log(O/H)_{\odot}}=8.69$ \citep{2009ARA&A..47..481A}, we found the local metallicity at the position of SN~2018bsz to be $\approx0.3-0.4\,Z_{\odot}$, in line with the general properties of SLSN host galaxies \citep[e.g.][]{2017MNRAS.470.3566C}.

\begin{figure}
    \centering
\includegraphics[width=0.8\textwidth]{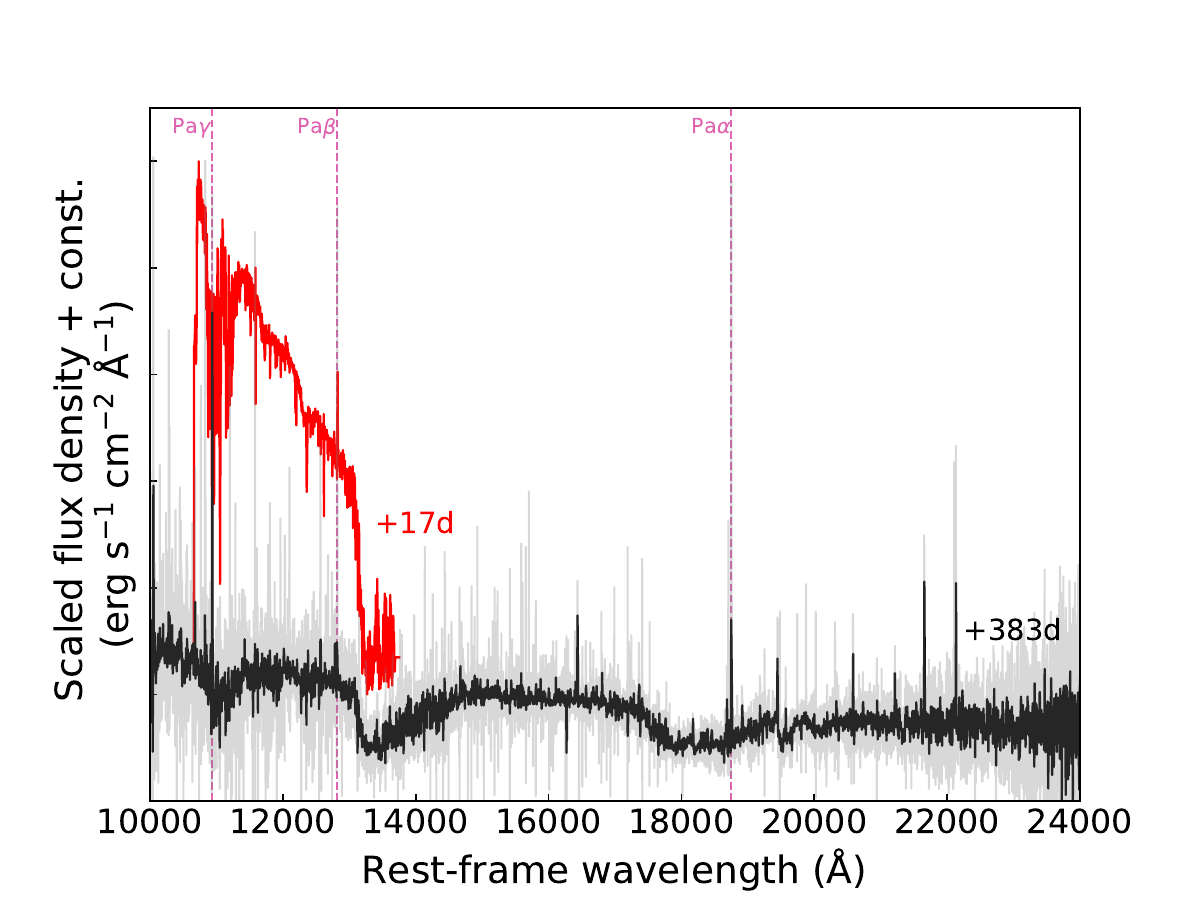}
    \caption{NIR spectra of SN~2018bsz. We extracted the SN position in the SINFONI data cube obtained at $+17$\,d. A narrow Pa$\beta$ is clearly detected, which is from the host H\,{\sc ii} region. No broad emission from the SN has been seen, confirming the hydrogen poor ejecta at peak brightness. The X-Shooter spectrum was taken on $+383$\,d, no trace from the SN been detected. The spectra shown in this figure are provided as Data behind the Figure.}
    \label{fig:SINFONI_XS}
\end{figure}

\section{Bolometric light-curve construction}
\label{sec:app_bol_cons}

\begin{figure}
    \centering
\includegraphics[width=0.8\textwidth]{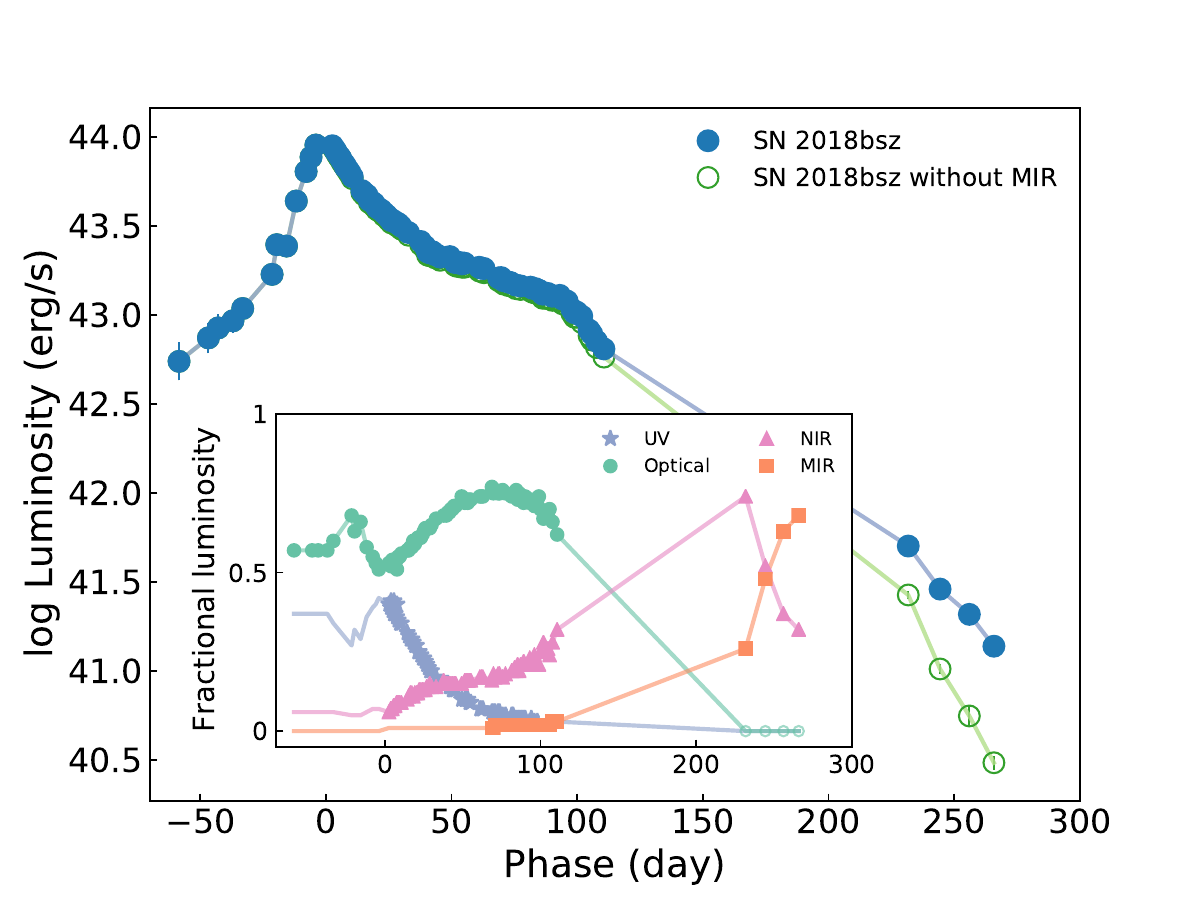}
    \caption{The bolometric light curve of SN~2018bsz. 
    The blue solid points show the lightcurve integrated over all bands as discussed in the text, while the empty green points show the integration excluding the MIR contribution. The inset panel shows the relative contribution of UV ($uvw2,uvm2,uvw1$), optical ($U,B,g,c,V,r,o,i,z$), NIR ($J,H,K_{s}$) and MIR ($W1,W2$).}
    \label{fig:bol_fra}
\end{figure}

\begin{figure}
    \centering
\includegraphics[width=0.8\textwidth]{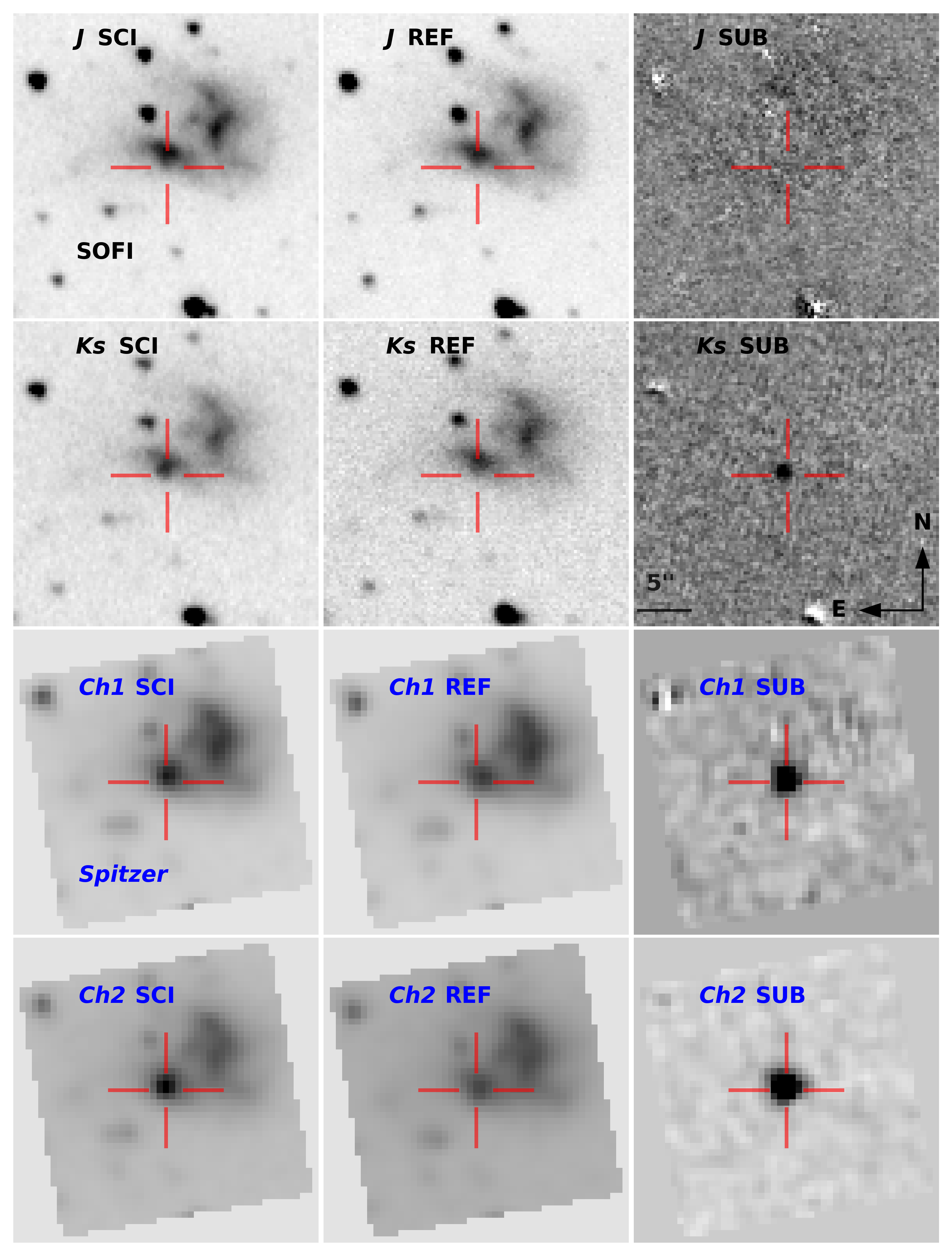}
\caption{SOFI and \textit{Spitzer} images of SN~2018bsz and template subtractions. The first row illustrates the SOFI $J$-band frame taken on $+266$\,d, the reference image taken on $+599$\,d, and the difference between the two (with a $3\sigma$ detection limit of $>21.92$\,mag). The second row shows SOFI $K_{\mathrm{s}}$-band images, as the same epoch as in the first row. The $K_{\mathrm{s}}$-band subtraction clearly detects a source at $+266$\,d where we detect the SN at $19.65\pm0.15$\,mag. The third row displays the \textit{Spitzer} $Ch1$ (3.6\,$\mu$m) frame taken on $+384$\,d, the reference image taken on $+565$\,d, and the result of their subtraction with a SN detection at $20.17\pm0.24$\,mag. The fourth row presents the \textit{Spitzer} $Ch2$ (4.5\,$\mu$m) images, at the same epoch as in the third row, where the SN was detected at $19.13\pm0.09$\,mag. Each panel has the same scale and orientation as marked in the second row.}
\label{fig:cutouts}
\end{figure}

We used the {\sc Superbol} code\footnote{https://github.com/mnicholl/superbol/blob/master/superbol.py} \citep{2018RNAAS...2..230N} to construct the bolometric light curve of SN~2018bsz. 
We integrated the flux for SN~2018bsz over the observed bands from UV to MIR between $+2$\,d and $+265$\,d, which gives very similar results to the bolometic lightcurve computed using blackbody extrapolations. 
We used the $K_{\mathrm{s}}$-band epochs as our reference passband and interpolated and extrapolated the other bands by fitting a low-order polynomial to estimate the magnitude at epochs with missing data in a given filter.
Unfortunately, during the pre-peak rising phase, we only have the optical light curve (with two colours), and thus we used the ATLAS $o$ band as the reference filter here. We assumed that the colours are the same as that at peak brightness to extrapolate the UV, other optical bands and NIR. 
For the MIR contribution, we smoothly interpolated the MIR magnitudes between the {\textit WISE} bands $+68$\,d and $+265$\,d. For earlier phases after peak, we run {\sc Superbol} again with no MIR data, and added a constant fraction to all epochs based on the MIR fraction at $+68$\,d of 1\%. Finally we assumed no MIR contribution before the peak.
We show the bolometric light curves with and without MIR contribution in Fig.\,\ref{fig:bol_fra}, those are basically the same for the first 150 days as the MIR contribution is negligible in the early time.

The inset in Figure\,\ref{fig:bol_fra} shows the fractional luminosity in different wavelength ranges for SN~2018bsz. 
During the peak luminosity, the fraction of flux in the UV ($uvw2, uvm2, uvw1$; 1597--4730\,\AA) is 40\%, optical ($U, B, g, c, V, r, o, i, z$; 3018--10647\,\AA) is 53\%, NIR ($J, H, K_{\mathrm{s}}$; 10707--24537\,\AA) is 6\% and MIR ($W1, W2$; 2.75--5.34\,$\mu$m) is only 1\% (assuming a constant fraction based on that at $+68$\,d).
After the peak, the UV fraction starts to decline, the optical fraction starts to increase and reaches 76\% around $+75$\,d, and the NIR fraction slowly increases. At $+40$\,d, the fraction from the UV and the NIR was equal, both contributing 15\% to the total luminosity. At $+111$\,d, the fraction of UV dropped to 3\%, the optical is 62\%, the NIR raised to 32\%, and the MIR is 3\%. 
At the late-time phase of $+232$\,d, there is no contribution from the UV and optical, the fraction from the NIR dominates at 74\%, and the MIR is 26\%. Just 34 days later the MIR has significantly increased to 68\%.

\begin{figure}
    \centering
\includegraphics[width=0.8\textwidth]{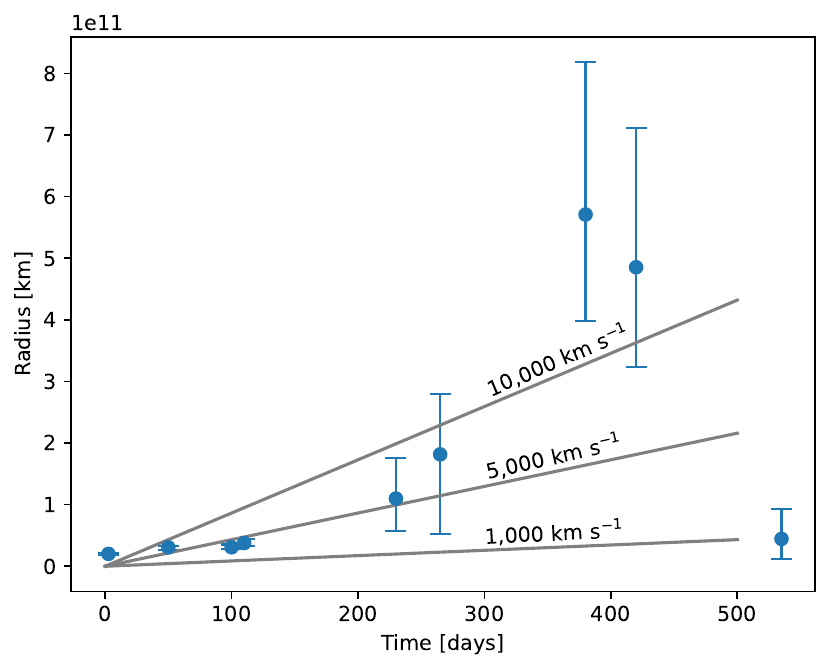}
    \caption{Evolution of the velocity as a function of blackbody radius and time for SN~2018bsz, based on the measurements listed in Table 1.}
    \label{fig:Velocity}
\end{figure}

\clearpage

\section{Observation log and photometry tables}
\label{sec:app_log}

\begin{deluxetable}{cccccccccc}
\tablecaption{\textit{Swift} UVOT aperture photometry for SN~2018bsz with host subtraction. Magnitudes are not corrected for Milky Way and host galaxy extinction, and all are in the AB system. The rest-frame phase is with respect to the {\textit r}-band peak epoch (MJD = 58267.5). A host template was constructed from images obtained between MJD = 58562 and 59068. 
\label{tab:phot_swift}}
\tablewidth{0pt}
\tabletypesize{\scriptsize}
\tablehead{
\colhead{Date} & 
\colhead{MJD} &
\colhead{Phase} &
\colhead{$m_{uvw2}$} &
\colhead{$m_{uvm2}$} &
\colhead{$m_{uvw1}$} &
\colhead{$m_{U}$} &
\colhead{$m_{B}$} &
\colhead{$m_{V}$} \\
\colhead{(UT)} &
\colhead{(d)} &
\colhead{(d)} &
\colhead{(mag)} &
\colhead{(mag)} &
\colhead{(mag)} &
\colhead{(mag)} &
\colhead{(mag)} &
\colhead{(mag)} 
}
\startdata
2018-05-31	&58269.517&1.96	      &18.04	(0.09)	&17.40	(0.06)	&16.67	(0.06)	&16.13	(0.06)	&16.08	(0.06)	&15.94	(0.08)\\
2018-06-03	&58272.723&5.09	      &18.08	(0.08)	&17.56	(0.06)	&16.78	(0.06)	&16.33	(0.06)	&16.26	(0.06)	&16.17	(0.08)\\
2018-06-06	&58275.260&7.56	      &18.22	(0.09)	&17.77	(0.06)	&17.02	(0.06)	&16.55	(0.06)	&16.43	(0.06)	&16.26	(0.09)\\
2018-06-12	&58281.665&13.80	      &18.72	(0.10)	&18.39	(0.08)	&17.51	(0.07)	&16.99	(0.07)	&16.76	(0.07)	&16.44	(0.09)\\
2018-06-15	&58284.252&16.32	      &18.95	(0.11)	&18.46	(0.10)	&17.76	(0.08)	&17.10	(0.07)	&16.88	(0.07)	&16.72	(0.10)\\
2018-06-18	&58287.640&19.62	      &-	&-	&18.00	(0.09)	&-	&-	&-\\
2018-06-21	&58290.759&22.65	      &19.37	(0.14)	&19.02	(0.12)	&18.09	(0.09)	&17.18	(0.07)	&16.89	(0.07)	&16.66	(0.10)\\
2018-06-24	&58293.215&25.05	      &19.64	(0.17)	&19.35	(0.15)	&18.16	(0.10)	&17.49	(0.08)	&17.02	(0.08)	&16.71	(0.11)\\
2018-06-27	&58296.736&28.48	      &19.53	(0.15)	&19.43	(0.16)	&18.39	(0.10)	&17.36	(0.07)	&16.92	(0.07)	&16.74	(0.10)\\
2018-07-06	&58305.679&37.19	      &20.17	(0.24)	&20.22	(0.21)	&18.87	(0.13)	&17.57	(0.09)	&17.16	(0.09)	&16.98	(0.14)\\
2018-07-10	&58309.316&40.73	      &20.37	(0.24)	&20.60	(0.25)	&19.19	(0.14)	&17.70	(0.08)	&17.26	(0.09)	&17.24	(0.14)\\
2018-07-12	&58311.511&42.87	      &21.17	(0.48)	&20.99	(0.39)	&19.13	(0.16)	&17.84	(0.11)	&17.53	(0.12)	&17.23	(0.18)\\
2018-07-15	&58314.396&45.68	      &21.40	(0.49)	&20.92	(0.30)	&19.30	(0.15)	&18.04	(0.10)	&17.47	(0.10)	&17.35	(0.15)\\
2018-07-18	&58317.187&48.40	      &21.37	(0.45)	&21.22	(0.35)	&19.56	(0.17)	&17.95	(0.09)	&17.44	(0.09)	&17.23	(0.13)\\
2018-07-20	&58319.189&50.34	      &21.19	(0.40)	&21.45	(0.40)	&19.51	(0.16)	&18.06	(0.09)	&17.46	(0.09)	&17.21	(0.13)\\
2018-07-24	&58323.672&54.71	      &21.16	(0.41)	&-	&19.38	(0.16)	&17.89	(0.09)	&17.36	(0.09)	&17.02	(0.12)\\
2018-08-04	&58334.891&65.64	      &21.62	(0.52)	&21.96	(0.56)	&19.55	(0.16)	&17.95	(0.09)	&17.44	(0.09)	&17.31	(0.13)\\
2018-08-10	&58340.975&71.56	      &21.63	(0.52)	&-	&19.78	(0.18)	&18.32	(0.11)	&17.56	(0.09)	&17.46	(0.14)\\
2018-08-21	&58351.757&82.07	      &-	&-	&20.45	(0.29)	&18.46	(0.12)	&17.74	(0.10)	&17.44	(0.14)\\
2018-08-25	&58355.442&85.66	      &-	&-	&20.59	(0.32)	&18.71	(0.13)	&17.93	(0.11)	&17.48	(0.14)\\
2018-08-29	&58359.360&89.47	      &-	&-	&20.61	(0.33)	&18.87	(0.15)	&18.05	(0.13)	&17.64	(0.17)\\
2018-09-02	&58363.251&93.26	      &-	&-	&20.63	(0.34)	&18.75	(0.14)	&17.87	(0.12)	&17.67	(0.18)\\
2018-09-07	&58368.792&98.66	      &-	&-	&20.77	(0.38)	&18.97	(0.17)	&18.23	(0.15)	&17.76	(0.19)\\
2018-09-17	&58378.294&107.91	      &-	&-	&-	&19.56	(0.25)	&18.67	(0.21)	&18.71	(0.38)\\
2018-09-29	&58390.281&119.59	      &-	&-	&-	&20.17	(0.38)	&-	&-\\
2018-10-04	&58395.424&124.60	      &-	&-	&-	&20.51	(0.50)	&-	&-\\
2018-10-10	&58401.104&130.13	      &-	&-	&-	&20.86	(0.62)	&20.19	(0.62)	&-\\
\enddata
\end{deluxetable}

\begin{longtable}{lllcccc}
\caption{Optical photometry for SN~2018bsz from GROND. The magnitudes are with host template subtraction, with 1$\sigma$ errors given in parentheses. Magnitudes are not corrected for Milky Way and host galaxy extinction, and are in the AB system. The rest-frame phase is with respect to the {\textit r}-band peak epoch (MJD = 58267.5).}
\label{tab:phot_grond_optical}
\\\hline
Date        & MJD       & Phase & $m_{g}$       & $m_{r}$       & $m_{i}$       & $m_{z}$\\
UT          &           & (d)   &(mag)          & (mag)         & (mag)         & (mag) \\ \hline 
\endfirsthead

\multicolumn{7}{c}%
{{\bfseries \tablename\ \thetable{} -- continued from previous page}} \\
\hline Date        & MJD       & Phase & $m_{g}$       & $m_{r}$       & $m_{i}$       & $m_{z}$\\
UT          &           & (d)   &(mag)          & (mag)         & (mag)         & (mag) \\\hline
\endhead

\multicolumn{7}{r}{{Continued on next page}} \\ \hline
\endfoot

\hline \hline
\endlastfoot

2018-06-01	&58270.139	&2.57	&15.89  (0.09)	&15.46  (0.14)	&15.37	(0.11)	&15.51	(0.07)\\
2018-06-02	&58271.229	&3.63	&15.90  (0.06)	&15.50  (0.15)	&15.41	(0.08)	&15.51	(0.07)\\
2018-06-03	&58272.284	&4.66	&15.97	(0.08)	&15.54	(0.14)	&15.46	(0.09)	&15.53	(0.08)\\
2018-06-04	&58273.202	&5.55	&16.04  (0.06)	&15.58  (0.13)  &15.47	(0.09)	&15.56	(0.08)\\
2018-06-05	&58274.229	&6.55	&16.09	(0.09)	&15.64	(0.13)	&15.52	(0.08)	&15.55	(0.08)\\
2018-06-06	&58275.202	&7.50	&16.17	(0.08)	&15.68	(0.13)	&15.53	(0.10)	&15.58	(0.07)\\
2018-06-07	&58276.181	&8.46	&16.22	(0.06)	&15.73	(0.13)	&15.58	(0.09)	&15.62	(0.07)\\
2018-06-08	&58277.188	&9.44	&16.27	(0.08)	&15.77	(0.13)	&15.63	(0.08)	&15.63	(0.08)\\
2018-06-09	&58278.258	&10.48	&16.37	(0.07)	&15.79	(0.16)	&15.68	(0.08)	&15.67	(0.07)\\
2018-06-13	&58282.168	&14.29	&16.53	(0.07)	&16.00	(0.13)	&15.85	(0.08)	&15.77  (0.07)\\
2018-06-14	&58283.088	&15.18	&16.55	(0.07)	&16.03	(0.14)	&15.89	(0.08)	&15.79	(0.08)\\
2018-06-15	&58284.241	&16.31	&16.59	(0.07)	&16.08	(0.15)	&15.77  (0.07)  &15.82	(0.07)\\
2018-06-16	&58285.246	&17.28	&16.72	(0.13)	&16.12	(0.13)	&16.00	(0.12)	&15.85	(0.09)\\
2018-06-17	&58286.132	&18.15	&16.48	(0.07)	&16.11	(0.13)	&15.96  (0.08)  &15.86	(0.07)\\
2018-06-18	&58287.160	&19.15	&16.64	(0.06)	&16.11	(0.13)	&15.98	(0.07)	&15.84	(0.07)\\
2018-06-19	&58288.211	&20.17	&16.66  (0.06)  &16.13  (0.12)	&16.00	(0.09)	&15.85	(0.08)\\
2018-06-20	&58289.219	&21.15	&16.61	(0.09)	&16.10	(0.16)	&16.01	(0.09)	&15.86	(0.07)\\
2018-06-21	&58290.070	&21.98	&16.65	(0.07)	&16.11	(0.14)	&16.02	(0.09)	&15.78	(0.06)\\
2018-06-22	&58291.208	&23.09	&16.68	(0.06)	&16.16	(0.13)	&16.04	(0.09)	&15.90	(0.07)\\
2018-06-23	&58292.275	&24.13	&16.66	(0.10)	&16.15	(0.13)	&16.05	(0.09)	&15.89	(0.07)\\
2018-06-24	&58293.135	&24.97	&16.70	(0.08)	&16.19	(0.14)	&16.05	(0.09)	&15.91	(0.08)\\
2018-06-25	&58294.144	&25.95	&16.68	(0.13)	&16.30	(0.13)	&16.07	(0.08)	&15.90	(0.09)\\
2018-06-26	&58295.071	&26.85	&16.70  (0.05)	&16.16  (0.14)	&16.08	(0.09)	&15.93	(0.07)\\
2018-06-27	&58296.161	&27.92	&16.72	(0.07)	&16.20	(0.13)	&16.12	(0.07)	&15.92	(0.07)\\
2018-06-28	&58297.106	&28.84	&16.74	(0.08)	&16.12  (0.13)	&16.13	(0.09)	&15.98	(0.07)\\
2018-06-29	&58298.192	&29.89	&16.72	(0.07)	&16.22  (0.13)	&16.13	(0.08)	&15.95	(0.07)\\
2018-07-02	&58301.203	&32.83	&16.77  (0.08)  &16.29  (0.13)  &16.18	(0.07)	&16.00	(0.08)\\
2018-07-07	&58306.184	&37.68	&16.89	(0.09)	&16.41	(0.13)	&16.33	(0.08)	&16.13	(0.08)\\
2018-07-08	&58307.961	&39.41	&16.96	(0.07)	&16.48	(0.12)	&16.41	(0.08)	&16.18	(0.09)\\
2018-07-10	&58309.005	&40.43	&17.16	(0.07)	&16.49	(0.12)	&16.42	(0.08)	&16.49  (0.08)\\
2018-07-11	&58310.051	&41.44	&17.11	(0.06)	&16.50	(0.13)	&16.64	(0.08)	&16.33  (0.07)\\
2018-07-12	&58311.036	&42.40	&17.04	(0.07)	&16.53	(0.13)	&16.49	(0.08)	&16.26	(0.08)\\
2018-07-13	&58312.019	&43.36	&17.09	(0.07)	&16.58	(0.13)	&16.54	(0.08)	&16.30  (0.06)\\
2018-07-14	&58313.109	&44.42	&17.09	(0.07)	&16.55	(0.15)	&16.51	(0.09)	&16.29	(0.07)\\
2018-07-15	&58314.130	&45.42	&17.09	(0.08)	&16.59	(0.12)	&16.57  (0.09)  &16.31	(0.07)\\
2018-07-19	&58318.105	&49.29	&17.11	(0.09)	&16.39  (0.13)	&16.42  (0.09)  &16.14  (0.12)\\
2018-07-21	&58320.200	&51.33	&17.13	(0.09)	&16.61	(0.14)	&16.60	(0.10)	&16.38	(0.07)\\
2018-07-22	&58321.082	&52.19	&17.12	(0.07)	&16.59	(0.13)	&16.58	(0.08)	&16.39	(0.07)\\
2018-07-23	&58322.127	&53.21	&17.11	(0.06)	&16.59	(0.14)	&16.58	(0.11)	&16.37	(0.08)\\
2018-07-24	&58323.024	&54.08	&17.12	(0.07)	&16.58	(0.13)	&16.57	(0.09)	&16.37	(0.07)\\
2018-07-25	&58324.115	&55.14	&17.08	(0.09)	&16.58	(0.12)	&16.56	(0.09)	&16.35	(0.08)\\
2018-07-31	&58330.098	&60.97	&17.11	(0.07)	&16.58	(0.14)	&16.57  (0.09)	&16.40	(0.08)\\
2018-08-01	&58331.043	&61.89	&17.12	(0.06)	&16.57	(0.15)	&16.61	(0.09)	&16.41	(0.08)\\
2018-08-02	&58332.079	&62.90	&17.12	(0.08)	&16.60	(0.15)	&16.62	(0.08)	&16.42  (0.07)\\
2018-08-04	&58334.962	&65.71	&17.34	(0.13)	&16.70	(0.12)	&16.73	(0.09)	&16.53	(0.09)\\
2018-08-08	&58338.020	&68.69	&17.19	(0.12)	&16.69	(0.13)	&16.73  (0.09)  &16.50	(0.08)\\
2018-08-09	&58339.043	&69.68	&17.29	(0.06)	&16.74	(0.13)	&16.76	(0.07)	&16.56	(0.07)\\
2018-08-10	&58340.096	&70.71	&17.33  (0.06)	&16.78  (0.12)  &16.78	(0.08)	&16.59	(0.07)\\
2018-08-12	&58342.057	&72.62	&17.38	(0.07)	&16.79	(0.13)	&16.82	(0.08)	&16.64	(0.07)\\
2018-08-13	&58343.057	&73.59	&17.40  (0.07)  &16.81  (0.13)  &16.82	(0.08)	&16.63	(0.07)\\
2018-08-15	&58345.043	&75.53	&17.40	(0.07)	&16.81	(0.13)	&16.83	(0.10)	&16.65	(0.07)\\
2018-08-16	&58346.978	&77.41	&17.43	(0.08)	&16.81	(0.13)	&16.83	(0.09)	&16.65	(0.08)\\
2018-08-21	&58351.040	&81.37	&17.47	(0.08)	&16.84	(0.13)	&16.83	(0.09)	&16.65	(0.08)\\
2018-08-21	&58351.990	&82.29	&17.45	(0.08)	&16.85	(0.14)	&16.83  (0.08)	&16.67	(0.08)\\
2018-08-22	&58352.995	&83.27	&17.47	(0.07)	&16.87	(0.13)	&16.85	(0.08)	&16.68	(0.07)\\
2018-08-24	&58354.008	&84.26	&-	            &-	            &-	            &-            \\
2018-08-24	&58354.997	&85.22	&17.52	(0.07)	&16.89	(0.13)	&16.91	(0.08)	&16.73	(0.07)\\
2018-08-26	&58356.005	&86.20	&17.53	(0.08)	&16.91	(0.13)	&16.92	(0.08)	&16.75  (0.09)\\
2018-08-27	&58357.006	&87.18	&17.57	(0.07)	&16.92	(0.13)	&16.91	(0.09)	&16.73	(0.07)\\
2018-08-28	&58358.054	&88.20	&17.55	(0.07)	&16.92  (0.08)	&16.93	(0.08)	&16.73  (0.07)\\
2018-08-28	&58358.972	&89.09	&17.57	(0.08)	&16.93	(0.12)	&16.93	(0.07)	&16.76	(0.07)\\
2018-08-29	&58359.973	&90.07	&17.54	(0.08)	&16.94	(0.14)	&16.90	(0.08)	&16.74	(0.08)\\
2018-09-02	&58363.006	&93.02	&17.53	(0.08)	&16.92	(0.13)	&16.92	(0.09)	&16.74  (0.07)\\
2018-09-03	&58364.004	&93.99	&17.59	(0.07)	&16.93	(0.14)	&16.94	(0.07)	&16.74	(0.08)\\
2018-09-05	&58366.024	&95.96	&17.63	(0.07)	&16.98	(0.13)	&16.98	(0.08)	&16.78	(0.08)\\
2018-09-06	&58367.996	&97.88	&17.72	(0.06)	&17.03	(0.15)	&17.05	(0.08)	&16.86	(0.07)\\
2018-09-07	&58368.997	&98.86	&17.76	(0.08)	&17.11	(0.13)	&17.06  (0.08)	&16.91	(0.09)\\
2018-09-08	&58369.997	&99.83	&17.83	(0.07)	&17.16	(0.13)	&17.14	(0.08)	&16.94	(0.07)\\
2018-09-10	&58371.997	&101.78	&17.96  (0.07)	&17.30  (0.13)  &17.28	(0.08)	&17.06	(0.07)\\
2018-09-13	&58374.994	&104.70	&18.13	(0.09)	&17.46	(0.14)	&17.49	(0.10)	&17.22	(0.08)\\
2018-09-15	&58376.045	&105.72	&18.13	(0.15)	&17.50	(0.15)	&17.50	(0.09)	&17.22	(0.07)\\
2018-09-16	&58377.998	&107.62	&18.39	(0.07)	&17.71	(0.14)	&17.71	(0.10)	&17.43	(0.07)\\
2018-09-20	&58381.040	&110.59	&18.56	(0.10)	&17.93	(0.12)	&17.96	(0.09)	&17.62	(0.07)\\
2019-01-22	&58505.353	&231.67	&$>21.46$		&$>23.47$		&$>20.77$		&$>22.08$  \\
2019-02-04	&58518.356	&244.33	&$>23.25$	    &$>23.34$		&$>21.61$		&$>21.32$	\\
2019-02-16	&58530.315	&255.98	&$>23.09$		&$>23.02$		&$>22.97$	    &$>22.60$  \\
2019-02-17	&58531.341	&256.98	&$>23.88$	    &$>23.26$	    &$>23.84$	    &$>22.44$  \\
2019-03-05	&58547.305	&272.53	&$>23.32$	    &$>23.22$       &$>23.01$	    &$>22.63$  \\
2019-04-04	&58577.349	&301.79	&$>23.16$	    &$>20.87$	    &$>22.72$	    &$>21.75$  \\
2019-04-07	&58580.325	&304.69	&$>23.38$       &$>23.30$	    &$>23.32$	    &$>22.82$         \\
2019-04-08	&58581.362	&305.70	&$>22.68$  	    &$>22.77$	    &$>22.67$	    &$>22.10$	 \\
2019-04-09	&58582.252	&306.57	&$>23.28$	    &$>23.91$	    &$>22.82$	    &$>21.36$	 \\
2019-04-10	&58583.408	&307.69	&$>22.87$	    &$>22.13$       &$>22.37$	    &$>21.68$  \\
2019-04-11	&58584.410	&308.67	&$>23.08$	    &$>22.95$	    &$>21.90$		&$>21.77$  \\
2019-04-13	&58586.402	&310.61	&$>23.40$	    &$>22.90$	    &$>22.32$	    &$>21.84$  \\
2019-06-05	&58639.149	&361.98	&$>23.53$	    &$>22.74$       &$>22.44$		&$>22.17$  \\
2019-08-05	&58700.043	&421.29	&$>23.70$	    &$>23.40$	    &$>23.31$		&$>22.96$ \\
2020-02-14	&58893.308	&609.53	& template	        & template	        & template         & template     \\
\hline
\end{longtable}

\clearpage

\newpage

\begin{longtable}{lllccc}
\caption{NIR photometry for SN~2018bsz with GROND. The magnitudes are with host template subtraction, with 1$\sigma$ errors given in parentheses. Magnitudes are given in AB mag. The rest-frame phase is with respect to the {\textit r}-band peak epoch (MJD = 58267.5). Note: we do not use the magnitudes with a $^{*}$ mark in our analysis.}
\label{tab:phot_grond_nir}
\\\hline
Date        & MJD       & Phase & $m_{J}$       & $m_{H}$       & $m_{Ks}$  \\
UT          &           & (d)   &(mag)          & (mag)         & (mag)     \\
\hline

\endfirsthead

\multicolumn{6}{c}%
{{\bfseries \tablename\ \thetable{} -- continued from previous page}} \\
\hline Date        & MJD       & Phase & $m_{J}$       & $m_{H}$       & $m_{Ks}$  \\
UT          &           & (d)   &(mag)          & (mag)         & (mag)     \\ \hline
\endhead

\multicolumn{6}{r}{{Continued on next page}} \\ 
\hline
\endfoot

\hline \hline
\endlastfoot

2018-06-01	&58270.139	&2.57	&15.68	(0.06)	&15.90	(0.06)	&16.05	(0.10)\\
2018-06-02	&58271.229	&3.63	&15.62	(0.04)	&15.81	(0.06)	&16.11	(0.08)\\
2018-06-03	&58272.284	&4.66	&15.63	(0.04)	&15.85	(0.06)	&16.09	(0.10)\\
2018-06-04	&58273.202	&5.55	&15.63	(0.06)	&15.73	(0.05)	&15.84	(0.09)\\
2018-06-05	&58274.229	&6.55	&15.70	(0.04)	&15.75	(0.08)	&15.93	(0.09)\\
2018-06-06	&58275.202	&7.50	&15.60	(0.07)	&15.74	(0.08)	&15.92	(0.09)\\
2018-06-07	&58276.181	&8.46  	&15.61	(0.09)	&15.82	(0.08)	&15.95	(0.11)\\
2018-06-08	&58277.188	&9.44	 &15.70	(0.08)	&15.83	(0.07)	&15.78	(0.11)\\
2018-06-09	&58278.258	&10.48	&15.73	(0.06)	&15.91	(0.10) &15.96	(0.11)\\
2018-06-13	&58282.168	&14.29	&15.78	(0.08)	&15.94 (0.10)	 &16.16	(0.08)\\
2018-06-14	&58283.088	&15.18	&15.78  (0.09)	&15.90	(0.10) &16.06	(0.09)\\
2018-06-15	&58284.241	&16.31	&15.55  (0.09)	&15.97 (0.08)	 & 16.16 (0.09)\\
2018-06-16	&58285.246	&17.28	&15.89	(0.07)	&15.92	(0.07) &16.07	(0.11)\\
2018-06-17	&58286.132	&18.15	&15.84  (0.11)	&16.07  (0.12)	& 16.17 (0.09)\\
2018-06-18	&58287.160	&19.15	&15.84	(0.10)	&15.99	(0.11)	&15.94	(0.09)\\
2018-06-19	&58288.211	&20.17	&15.93	(0.07)	&15.99  (0.08)	&16.25	(0.09)\\
2018-06-20	&58289.219	&21.15	&15.89	(0.05)	&16.03	(0.06)	&16.26	(0.09)\\
2018-06-21	&58290.070	&21.98	&15.85	(0.10)	&15.94	(0.12)	&16.29	(0.08)\\
2018-06-22	&58291.208	&23.09	&15.90	(0.11)	&16.03	(0.11)	&16.02	(0.10)\\
2018-06-23	&58292.275	&24.13	&15.85	(0.08)	&16.02  (0.09)  &16.25	(0.10)\\
2018-06-24	&58293.135	&24.97	&15.94	(0.05)	&16.13  (0.06)	&16.12	(0.10)\\
2018-06-25	&58294.144	&25.95	&15.96	(0.04)	&16.14	(0.06)	&16.40	(0.10)\\
2018-06-26	&58295.071	&26.85	&15.90	(0.07)	&16.08  (0.08)	&16.27	(0.09)\\
2018-06-27	&58296.161	&27.92	&15.88	(0.12)	&16.15	(0.12)	&16.23	(0.09)\\
2018-06-28	&58297.106	&28.84	&15.94	(0.06)	&15.98	(0.08)	&16.22	(0.09)\\
2018-06-29	&58298.192	&29.89	&15.93	(0.08)	&16.15	(0.08)	&16.29	(0.10)\\
2018-07-02	&58301.203	&32.83	&16.02	(0.05)	&16.19	(0.06)	&16.38	(0.10)\\
2018-07-07	&58306.184	&37.68	&16.02	(0.05)	&16.31	(0.07)	&16.35	(0.10)\\
2018-07-08	&58307.961	&39.41	&16.20	(0.06)	&16.25	(0.07)	&16.47	(0.10)\\
2018-07-10	&58309.005	&40.43	&16.40	(0.10)	&16.33	(0.14)	&16.52	(0.09)\\
2018-07-11	&58310.051	&41.44	&16.29	(0.08)	&16.44  (0.12)	&16.60	(0.10)\\
2018-07-12	&58311.036	&42.40	&16.17	(0.08)	&16.63	(0.13)	&16.47	(0.08)\\
2018-07-13	&58312.019	&43.36	&16.37	(0.08)	&16.43	(0.09)	&16.52	(0.09)\\
2018-07-14	&58313.109	&44.42	&16.36	(0.11)	&16.49  (0.16)	&16.55	(0.10)\\
2018-07-15	&58314.130	&45.42	&16.36  (0.07)	&16.48  (0.07)	&16.62  (0.09)\\
2018-07-19	&58318.105	&49.29	&16.29	(0.05)	&16.52	(0.06)	&16.83	(0.13)\\
2018-07-21	&58320.200	&51.33	&16.29	(0.04)	&16.46	(0.07)	&16.74	(0.10)\\
2018-07-22	&58321.082	&52.19	&16.34	(0.06)	&16.60	(0.07)	&16.56	(0.09)\\
2018-07-23	&58322.127	&53.21	&16.31	(0.07)	&16.48	(0.07)	&16.53	(0.10)\\
2018-07-24	&58323.024	&54.08	&16.38	(0.06)	&16.56	(0.08)	&16.57	(0.09)\\
2018-07-25	&58324.115	&55.14	&16.39	(0.04)	&16.57	(0.06)	&16.39	(0.11)\\
2018-07-31	&58330.098	&60.97	&16.41  (0.09)	&16.45  (0.07)	&16.53  (0.10)\\
2018-08-01	&58331.043	&61.89	&16.34	(0.07)	&16.53	(0.08)	&16.63	(0.09)\\
2018-08-02	&58332.079	&62.90	&16.35	(0.06)	&16.59	(0.07)	&16.45	(0.09)\\
2018-08-04	&58334.962	&65.71	&16.63	(0.09)	&16.60	(0.10)	&16.00$^{*}$  (0.29)\\
2018-08-08	&58338.020	&68.69	&16.62  (0.04)	&16.65  (0.05)	&16.98  (0.11)\\
2018-08-09	&58339.043	&69.68	&16.49	(0.08)	&16.59	(0.08)	&16.51	(0.09)\\
2018-08-10	&58340.096	&70.71	&16.58	(0.06)	&16.65  (0.07)	&16.79	(0.10)\\
2018-08-12	&58342.057	&72.62	&16.51	(0.08)	&16.65	(0.09)	&16.72	(0.09)\\
2018-08-13	&58343.057	&73.59	&16.52	(0.05)	&16.56  (0.07)  &16.77	(0.09)\\
2018-08-15	&58345.043	&75.53	&16.69	(0.10)	&16.61	(0.08)	&16.73	(0.10)\\
2018-08-16	&58346.978	&77.41	&16.78	(0.12)	&16.43  (0.11)  &16.75	(0.08)\\
2018-08-21	&58351.040	&81.37	&16.51	(0.08)	&16.64	(0.09)	&16.58	(0.08)\\
2018-08-21	&58351.990	&82.29	&16.58  (0.06)	&16.61  (0.07)	&16.63  (0.08)\\
2018-08-22	&58352.995	&83.27	&16.57	(0.06)	&16.62  (0.08)	&16.43	(0.09)\\
2018-08-24	&58354.008	&84.26	&16.64	(0.04)	&16.63  (0.04)	&16.72  (0.11)\\
2018-08-24	&58354.997	&85.22	&16.57	(0.06)	&16.52	(0.07)	&16.51	(0.09)\\
2018-08-26	&58356.005	&86.20	&16.71	(0.07)	&16.67  (0.08)	&16.57	(0.10)\\
2018-08-27	&58357.006	&87.18	&16.72	(0.07)	&16.52	(0.07)	&16.55	(0.09)\\
2018-08-28	&58358.054	&88.20	&16.60	(0.07)	&16.62	(0.07)	&16.34	(0.09)\\
2018-08-28	&58358.972	&89.09	&16.55	(0.06)	&16.53	(0.07)	&16.54	(0.08)\\
2018-08-29	&58359.973	&90.07	&16.65	(0.04)	&16.58	(0.07)	&16.64	(0.08)\\
2018-09-02	&58363.006	&93.02	&16.53	(0.09)	&16.54	(0.09)	&16.29	(0.09)\\
2018-09-03	&58364.004	&93.99	&16.62	(0.07)	&16.67  (0.08)	&16.44	(0.10)\\
2018-09-05	&58366.024	&95.96	&16.63	(0.06)	&16.57  (0.08)  &16.40	(0.10)\\
2018-09-06	&58367.996	&97.88	&16.86	(0.07)	&16.67  (0.11)	&16.82	(0.10)\\
2018-09-07	&58368.997	&98.86	&16.95  (0.08)	&16.77  (0.12)	&16.81  (0.10)\\
2018-09-08	&58369.997	&99.83	&16.81	(0.06)	&16.59	(0.07)	&16.59	(0.09)\\
2018-09-10	&58371.997	&101.78	&16.85	(0.05)	&16.53	(0.07)	&16.35	(0.10)\\
2018-09-13	&58374.994	&104.70	&17.02	(0.06)	&16.82	(0.06)	&16.77	(0.10)\\
2018-09-15	&58376.045	&105.72	&17.36	(0.05)	&16.85	(0.05)	&16.72	(0.10)\\
2018-09-16	&58377.998	&107.62	&17.13	(0.07)	&16.87	(0.09)	&16.77	(0.08)\\
2018-09-20	&58381.040	&110.59	&17.18	(0.06)	&16.87	(0.07)	&16.59	(0.09)\\

2019-01-22	&58505.353	&231.67	&19.88	(0.15)	&18.85	(0.14)	&18.03	(0.14)\\
2019-02-04	&58518.356	&244.33	&21.48  (0.28)	&20.14  (0.17)	&18.59	(0.11)\\
2019-02-16	&58530.315	&255.98	&$>21.43$	    &20.60  (0.26)	&19.27  (0.26)\\
2019-02-17	&58531.341	&256.98	&$>20.44$		&20.78  (0.24)	&$>18.59$\\
2019-03-05	&58547.305	&272.53	&$>21.59$	    &$>20.37$	    &$>19.55$\\
2019-04-04	&58577.349	&301.79	&$>19.52$		&$>19.51$       &$>19.29$\\
2019-04-07	&58580.325	&304.69	&$>21.35$	    &$>20.23$		&$>19.40$\\
2019-04-08	&58581.362	&305.70	&$>21.12$		&$>20.22$		&$>19.29$\\
2019-04-09	&58582.252	&306.57	&$>20.37$	    &$>20.67$	    &$>19.13$\\
2019-04-10	&58583.408	&307.69	&$>21.95$		&$>20.58$		&$>19.50$\\
2019-04-11	&58584.410	&308.67	&$>21.84$		&$>20.49$	    &$>19.30$\\
2019-04-13	&58586.402	&310.61	&$>21.22$		&$>20.49$	    &$>19.31$\\
2019-06-05	&58639.149	&361.98	&$>20.56$		&$>20.16$		&$>18.13$\\
2019-08-05	&58700.043	&421.29	&$>21.56$		&$>20.92$	    &$>19.75$\\
2020-02-14	&58893.308	&609.53	&template			&template		    &template	\\
\hline
\end{longtable}

\begin{deluxetable}{cccccc}
\tablecaption{SOFI observations for SN~2018bsz. The magnitudes are after host template subtraction and given in AB mag. \label{tab:phot_sofi}}
\tablewidth{0pt}
\tabletypesize{\scriptsize}
\tablehead{
\colhead{Date} & 
\colhead{MJD} &
\colhead{Phase} &
\colhead{$m_{J}$} &
\colhead{$m_{K_{s}}$} \\ 
\colhead{(UT)} &
\colhead{(d)} &
\colhead{(d)} &
\colhead{(mag)} &
\colhead{(mag)} &
}
\startdata
2019-02-26  & 58540.344 & 265.75 & $>21.92$ & 19.65(0.15)  \\
2019-05-02  & 58605.069 & 328.79 & - & $>20.67$  \\
2019-07-31  & 58695.105 & 416.49 & - & $>21.11$  \\
2020-02-03  & 58882.373 & 598.88 & template & -  \\
2020-02-04  & 58883.338 & 599.82 & - & template  \\
\enddata
\end{deluxetable}

\begin{deluxetable}{cccccc}
\tablecaption{\textit{Spitzer} MIR aperture photometry for SN~2018bsz with host subtraction. We use the 2019-12-30 epoch as the host template assuming the SN flux is negligible. The 3-$\sigma$ uncertainty is estimated by the standard deviation of the sky background near the SN position. The {\textit Ch1} and {\textit Ch2} channels correspond to 3.6~$\mu$m and 4.5~$\mu$m, respectively. The magnitudes are reported in the AB system. \label{tab:phot_spitzer}}
\tablewidth{0pt}
\tabletypesize{\scriptsize}
\tablehead{
\colhead{Date} & 
\colhead{MJD} &
\colhead{Phase} &
\colhead{$m_{Ch1}$} &
\colhead{$m_{Ch2}$} \\
\colhead{(UT)} &
\colhead{(d)} &
\colhead{(d)} &
\colhead{(mag)} &
\colhead{(mag)} &
}
\startdata
2019-06-27  & 58661.648 & 383.90 & 20.17(0.24) & 19.13(0.09)  \\
2019-07-17  & 58681.576 & 403.31 & 20.46(0.23) & 19.42(0.11)  \\
2019-11-29  & 58816.551 & 534.77 & 21.84$^{*}$(0.70) &  21.59$^{*}$(0.55) \\
2019-12-30  & 58847.113 & 564.54 & template & template \\
\enddata
\tablecomments{{\it \bf $^{*}$}We note that the magnitudes at $+535$\,d could be possibly over-subtracted from the template as there are only 30 days in between. Therefore we assume a linear decay since $+384$\,d and extrapolate the magnitudes at $+535$\,d. These extrapolations yield 22.42 and 21.41\,mag in $m_{Ch1}$ and $m_{Ch2}$, respectively, which are consistent with the template-subtracted magnitudes within the uncertainties.}
\end{deluxetable}

\begin{deluxetable}{cccccc}
\tablecaption{NEOWISE MIR light curves for SN~2018bsz with host-galaxy subtraction. The magnitudes of the host galaxy are estimated from the mean magnitude of all NEOWISE epochs up until the SN explosion. The 3-$\sigma$ uncertainties include statistic and systematic errors. The {\textit W1} and {\textit W2} bands correspond to 3.4~$\mu$m and 4.6~$\mu$m, respectively. The magnitudes are reported in the AB system. \label{tab:phot_neowise}}
\tablewidth{0pt}
\tabletypesize{\scriptsize}
\tablehead{
\colhead{Date} & 
\colhead{MJD} &
\colhead{Phase} &
\colhead{$m_{W1}$} &
\colhead{$m_{W2}$} \\ 
\colhead{(UT)} &
\colhead{(d)} &
\colhead{(d)} &
\colhead{(mag)} &
\colhead{(mag)} &
}
\startdata
2018-08-07  & 58337.198 & 67.89 & 17.60(0.14) & 17.83(0.16)  \\
2019-02-25  & 58539.742 & 265.16 & 18.88(0.42) & 18.01(0.19)  \\
2019-08-06  & 58701.536 & 422.75 & - & 19.64(0.85)  \\
Host galaxy & - & & 16.20(0.01) & 16.65(0.02) \\
\enddata
\tablecomments{{\it \bf a}: average MJD of all 12-18 exposures in a single epoch.} 
\end{deluxetable}

\begin{deluxetable}{cccccccccccc}
\tablecaption{Spectroscopic observations. 
\label{tab:log_spec}}
\tablewidth{0pt}
\tabletypesize{\scriptsize}
\tablehead{
\colhead{Date} & 
\colhead{MJD} &
\colhead{Phase} & 
\colhead{Telescope} &
\colhead{Instrument} &
\colhead{Grism or Grating} &
\colhead{Exp. time} &
\colhead{Slit} &
\colhead{Resolution} &
\colhead{Range} & \\
\colhead{} & 
\colhead{} & 
\colhead{(day)} &
\colhead{} & 
\colhead{} & 
\colhead{} & 
\colhead{(s)} &
\colhead{(\arcsec)} &
\colhead{(\AA)} &
\colhead{(\AA)} & 
}
\startdata
2018-05-23 & 58261.290 & $-6$ & NTT & EFOSC2 & Gr\#11/Gr\#16 & 1500/1500 & 1.0/1.0 & 13.8/13.4 & 3400-7400/6000-9900$^{a}$  \\
2019-06-16 & 58285 & $+17$ & VLT & SINFONI & J & 6300 & IFU &  & 11,000-14,000 \\
2018-08-16 & 58346.561 & $+77$ & ANU 2.3-m & WiFeS & B3000/R3000 & 1800/1800 & IFU & 1.6/2.5 & 3500-5700/5400-9500 \\
2018-09-17 & 58378.035 & $+108$ & NTT & EFOSC2 & Gr\#11/Gr\#16 & 2700/2700 & 1.0/1.0 & 13.8/13.4 & 3400-7400/6000-9900 \\
2019-04-30 & 58603 & $+327$ & VLT & X-Shooter & UVB/VIS/NIR & 3600/3712/3600 & 0.9/0.9/1.0 & 1/1.1/3.3 & 3000-5560/5450-10,200/10,000-20,600 \\
2019-05-05 & 58608 & $+332$ & VLT & X-Shooter & UVB/VIS/NIR & 7200/7424/7200 & 0.9/0.9/1.0 & 1/1.1/3.3 & 3000-5560/5450-10,200/10,000-20,600 \\
2019-05-06 & 58609 & $+333$ & VLT & X-Shooter & UVB/VIS/NIR & 3600/3712/3600 & 0.9/0.9/1.0 & 1/1.1/3.3 & 3000-5560/5450-10,200/10,000-20,600 \\
2019-06-27 & 58661 & $+383$ & VLT & X-Shooter & UVB/VIS/NIR & 7200/7424/7200 & 0.9/0.9/1.0 & 1/1.1/3.3 & 3000-5560/5450-10,200/10,000-20,600 \\
2019-08-01 & 58696 & $+417$ & VLT & MUSE &  & 2139 & IFU &  & 4650-9300 \\
\enddata
\tablecomments{{\it \bf a}: The $-6$\,d spectrum is taken from Anderson et al. (2018). {\it \bf b}: The $+108$\,d spectrum has been presented in Pursiainen et al. (2022). }
\end{deluxetable}

\bibliography{janet.bib}
\bibliographystyle{apj}

\end{document}